\documentclass[aps,prd,twocolumn,groupedaddress,floatfix,longbibliography,superscriptaddress]{revtex4-2}

\usepackage{amsmath}
\usepackage{amssymb}
\usepackage{amsfonts}
\usepackage{bbold}
\usepackage{bm}
\usepackage{cancel}
\usepackage{times,float}
\usepackage{graphicx}
\graphicspath{{Figures/}}
\usepackage[usenames,dvipsnames,svgnames]{xcolor}
\usepackage{hyperref}
\hypersetup{colorlinks=true, linkcolor=Blue, citecolor=Blue,urlcolor=Blue}
\usepackage{multirow}
\usepackage{ulem}
\usepackage{color,epsfig}
\usepackage{slashed}
\usepackage{feynmf}
\usepackage{array}
\usepackage{physics}
\usepackage{subcaption}
\usepackage{soul}

\begin{document}

\title{Disorder-induced modulation of the nonlinear Hall effect in Weyl semimetals}

\author{Juan A. Ca\~nas}
\email{juan.canas@correo.nucleares.unam.mx}
\address{Instituto de Ciencias Nucleares, Universidad Nacional Aut\'{o}noma de M\'{e}xico, 04510 Ciudad de M\'{e}xico, M\'{e}xico}

\author{Daniel A. Bonilla}
\email{daniel.bonillam@correo.nucleares.unam.mx}
\address{Instituto de Ciencias Nucleares, Universidad Nacional Aut\'{o}noma de M\'{e}xico, 04510 Ciudad de M\'{e}xico, M\'{e}xico}

\author{A. Mart\'{i}n-Ruiz}
\email{alberto.martin@nucleares.unam.mx}
\address{Instituto de Ciencias Nucleares, Universidad Nacional Aut\'{o}noma de M\'{e}xico, 04510 Ciudad de M\'{e}xico, M\'{e}xico}

\begin{abstract}
We study the effects of impurity scattering on the nonlinear Hall response of Weyl semimetals within the semiclassical Boltzmann approach. We derive the rectified and second-harmonic conductivity tensors for a general momentum-dependent transport relaxation time and evaluate this quantity microscopically for short-range, Gaussian, screened Coulomb, and magnetic impurities. For scalar disorder, the relaxation time is isotropic and chirality independent. The nonlinear Hall response is then determined by the anomalous velocity associated with the Berry curvature, and a finite net response requires energetically inequivalent Weyl nodes to avoid cancellation between opposite chiralities. Polarized magnetic impurities lead to a qualitatively different behavior. We find that interference between the first- and second-order Born amplitudes generates a helicity-dependent anisotropic correction to the relaxation time. This anisotropy changes the tensor structure of the nonlinear Hall conductivity and gives rise to a finite second-harmonic current for suitable orientations of the electric field relative to the impurity polarization. For representative parameters, however, the anisotropic contribution is several orders of magnitude smaller than the dominant isotropic response. These results establish how the microscopic form of impurity scattering enters the nonlinear Hall response of Weyl semimetals through the transport relaxation time.
\end{abstract}

\maketitle


\section{Introduction}

Topological semimetals have garnered significant interest in contemporary condensed matter physics owing to their unconventional electronic structures and the robust transport phenomena arising from nontrivial band topology \cite{LIU2024101343,YuAFM2025,Bernevig2022,Schoop2020,Deng2019,Zou2019,Bernevig2018}. Within this broad family, Weyl semimetals occupy a central position as three-dimensional systems in which nondegenerate conduction and valence bands touch at isolated points, known as Weyl nodes, in the Brillouin zone. These nodes act as monopoles of Berry curvature and occur in sets with vanishing total chirality, while their realization requires the breaking of either spatial-inversion ($P$) or time-reversal ($T$) symmetry \cite{ZhongAM2025,JinPRL2020,Wang_PRB2019,YanARCMP2017,Young_PRL2012}. Their nontrivial topology also gives rise to surface Fermi arcs connecting the surface projections of Weyl nodes with opposite chirality. Weyl semimetals have been identified experimentally through spectroscopic measurements, including the observation of Fermi arcs by angle-resolved photoemission spectroscopy \cite{CHENM2020,LvNRP2019,Deng2019,DengNP2016,XuS2015,WengPRX2015,NeupaneNC2014}, and through transport signatures associated with their band topology \cite{YangP2025}, such as the chiral anomaly and negative longitudinal magnetoresistance \cite{LiangPRX2018,HuangPRX2015}, Shubnikov--de Haas oscillations \cite{HuangPRX2015,HePRL2014}, and the quantum Hall effect \cite{ZhangN2019,UchidaNC2017}. The effects of torsional strain and dislocations on electronic and thermoelectric transport have also been studied for a single defect under an external magnetic field \cite{BonillaNano2021} and for a dilute uniform distribution of defects \cite{BonillaNano2022,BonillaNA2024}. In addition, the combined effects of magnetic fields and torsional strain on electron--phonon interactions and Peierls-type lattice instabilities have been analyzed in the quantum limit \cite{ParraPRB2026}. These developments have opened the possibility of exploring topological transport beyond the linear-response regime.

Beyond their extensively investigated linear properties, topological semimetals display a rich variety of nonlinear electronic and optical phenomena associated with their band geometry, low-energy quasiparticle structure, and nonequilibrium carrier dynamics \cite{ShenPQE2024,WangAPS2023,ZuberFoP2021,WangPRB2020,GorbarLTP2018}. Such effects have been investigated across different classes of topological semimetals, including Weyl, Dirac, and nodal-line systems. Particularly important examples are second-order responses such as photogalvanic and rectification effects \cite{Xin_AcPS2021,deJuan_NatC2017,Ma_NatPhys2017}, electrochemical transport \cite{PhysRevB.103.035102}, planar Hall effects \cite{PhysRevB.108.155132, sci_reports_10.1038}, and high-harmonic generation \cite{Liu_PRB2025,Lv_NatC2021,Wu_NatPhys2017}. Within the conventional electric-dipole description, a bulk second-order current is forbidden by spatial-inversion symmetry \cite{Zhang_NatP2025,BOYD20081}. Nevertheless, recent studies have shown that finite nonlinear responses may also arise in nominally centrosymmetric systems through nonequilibrium driving, additional spatial structure, or disorder-induced scattering mechanisms, without requiring explicit inversion breaking in the equilibrium band structure \cite{Suo_ACSP2026,Salawu_NatC2025,Du_NatC2019}. These results demonstrate that nonlinear transport is controlled not only by the symmetries and geometric properties of the electronic bands, but also by the microscopic processes governing the nonequilibrium distribution of charge carriers.

To capture these nonequilibrium effects, recent theoretical works have rigorously extended the semiclassical Boltzmann framework to incorporate extrinsic nonlinear Hall mechanisms, such as skew-scattering and side-jump processes \cite{Ma_PRB2025,Du_NatRP2021,Sinitsyn_2008}, and have introduced modified semiclassical equations to account for higher-order quantum corrections \cite{Xiao_PRB2019}. Taking a complementary route, in the present work we investigate how the explicit microscopic structure of impurity scattering combines with symmetry breaking to determine the second-order Hall response of a Weyl semimetal. We consider an isotropic two-node model in which nodes of opposite chirality may be displaced relative to each other in energy, thereby producing different local Fermi energies and preventing the exact cancellation of their Berry-curvature contributions. Within the semiclassical Boltzmann formalism, rather than incorporating anomalous driving terms into the equations of motion, we derive the rectified and second-harmonic conductivity tensors by focusing exclusively on a general momentum-dependent transport relaxation time. We then determine this relaxation time microscopically for short-range, finite-range Gaussian, screened Coulomb, and magnetic exchange impurities. This approach allows us to separate the geometric contribution associated with the Berry curvature from the energy and angular dependence introduced by each scattering mechanism. In particular, while scalar disorder produces an isotropic relaxation time, polarized magnetic impurities introduce a preferred direction and generate an anisotropic tensorial structure. The Weyl-node chirality then provides the natural framework for analyzing how impurity scattering and nodal inequivalence jointly control the magnitude, frequency dependence, and symmetry of the nonlinear Hall response.

The remainder of this article is organized as follows. In Section \ref{Sec:Semiclassical_formalism}, we introduce the semiclassical formalism for nonlinear Hall effects, solve the Boltzmann equation up to second order in the external field, and classify the resulting contributions into rectified quasiparticle (DC) responses and second harmonic (AC) responses. Section \ref{sec_relaxation_times} presents the low-energy model of a two-node Weyl semimetal and the calculation of the momentum-dependent relaxation time for different types of disorder potentials, including short-range, Gaussian, Coulomb and magnetic impurities. In Section \ref{Nonlinear_Hall_WSM_Section}, we evaluate the nonlinear Hall conductivities, analyzing the disorder-induced effects on both the rectification and second harmonic response tensors. Finally, Section \ref{Conclusion_Section} summarizes our main conclusions.

\section{Semiclassical formalism}
\label{Sec:Semiclassical_formalism}

In this section we establish the general semiclassical framework used
throughout this work to describe nonlinear Hall transport in Weyl
semimetals. Our goal is to derive model-independent expressions for the
second-order conductivity tensors within the Boltzmann formalism,
keeping the treatment as general as possible. At this stage, no
particular microscopic scattering mechanism is specified. Instead, all
the effects of disorder are encoded in the transport relaxation time,
which will be evaluated for different impurity models in the following
section.

The dynamics of Bloch electrons in topological materials is governed by the semiclassical equations of motion, which incorporate the effects of the Berry curvature in addition to the ordinary band dispersion \cite{PhysRevB.59.14915, RevModPhys.82.1959}. Within a given energy band and Weyl node, it is convenient to introduce the collective index $\alpha=(s,\chi)$, where $s=\pm1$ labels the conduction
and valence bands, while $\chi=\pm1$ denotes the chirality of the Weyl
node. The corresponding band dispersion and Berry curvature are denoted
by $E _{\alpha\mathbf{k}}$ and $\boldsymbol{\Omega}_{\alpha\mathbf{k}}$,
respectively.

In the presence of external electric and magnetic fields, the
semiclassical equations of motion take the form
\begin{align}
    \dot{\mathbf{r}} _{\alpha} &= \frac{1}{\hbar} \nabla _{\mathbf{k}} \, E _{\alpha \mathbf{k}} - \dot{\mathbf{k}} _{\alpha} \times \boldsymbol{\Omega} _{\alpha \mathbf{k}} \\ \hbar \dot{\mathbf{k}} _{\alpha} &= - e  \mathbf{E}  - e \dot{\mathbf{r}} _{\alpha} \times \mathbf{B} .  
\end{align}
These equations describe the coupled evolution of the center of a Bloch
wave packet in real and momentum space. The Berry-curvature term
produces an anomalous velocity, which is responsible for a variety of
topological transport phenomena, including anomalous Hall and nonlinear
Hall effects.

The nonequilibrium dynamics of the electronic distribution is described by the semiclassical Boltzmann equation,
\begin{align}
\frac{\partial f_{\alpha}}{\partial t}+\dot{r}_{\alpha}\cdot\nabla_{r}f_{\alpha}+\dot{k}_{\alpha}\cdot\nabla_{k}f_{\alpha}=\left(\frac{\partial f_{\alpha}}{\partial t}\right)_{coll},
\end{align}
where $f_{\alpha}(r,k,t)$ denotes the occupation probability of a Bloch state. The drift terms account for the action of the external fields, whereas the collision integral incorporates the effect of impurity scattering. Rather than expanding this collision term to explicitly extract extrinsic side-jump and skew-scattering anomalous driving forces \cite{Ma_PRB2025, Du_NatRP2021,Sinitsyn_2008}, our approach captures the fundamental effects of disorder strictly through the momentum-dependent transport relaxation time.

Throughout this work we consider spatially homogeneous systems driven by
an oscillating electric field,
\begin{align}
    \mathbf{E} (t) = \frac{1}{2} \left( \boldsymbol{\mathcal{E}} e ^{i \omega t} + \boldsymbol{\mathcal{E}} ^{\ast} e ^{- i \omega t} \right)
\end{align}
and neglect magnetic-field-induced orbital motion, since our primary interest is the nonlinear Hall response generated by the electric field. Within the relaxation-time approximation, the collision integral is written as
\begin{equation}
\left(
\frac{\partial f _{\alpha} (\mathbf{k} , t )}{\partial t}
\right) _{\rm{coll}} = - \frac{\delta f _{\alpha} (\mathbf{k} , t )}{\tau _{\alpha \mathbf{k} }} ,
\end{equation}
where $\tau _{\alpha \mathbf{k}}$ is the transport relaxation time. At this point we leave its microscopic form unspecified, emphasizing that the following derivation applies to an arbitrary momentum-dependent relaxation time.

The Boltzmann equation therefore becomes
\begin{align}
     \frac{\partial f _{\alpha} (\mathbf{k} , t ) }{\partial t} - \frac{e}{\hbar} \, \mathbf{E} (t) \cdot \nabla _{\mathbf{k}} \, f _{\alpha} (\mathbf{k} , t ) = - \frac{\delta f _{\alpha} (\mathbf{k} , t )}{\tau _{ \alpha \mathbf{k} }  }  .    \label{eq_Boltzmann}
\end{align}
The nonlinear Hall response originates from the nonequilibrium correction
to the electronic distribution induced by the oscillating electric field.
To obtain the response up to second order in the electric field, we expand the distribution function around equilibrium as \cite{Morimoto_PRB2016}
\begin{equation}
f _{\alpha} (\mathbf{k},t) = f _{0} ( E _{\alpha \mathbf{k}}) + \delta f _{\alpha} (\mathbf{k},t),
\end{equation}
where $f_{0} (E)$ denotes the Fermi-Dirac distribution and $\delta f _{\alpha}$ is generated by the external field. The nonequilibrium correction is further expanded as
\begin{equation}
\delta f _{\alpha}  = \delta f _{\alpha} ^{(1)} + \delta f _{\alpha} ^{(2)} + \cdots ,
\end{equation}
where $\delta f _{\alpha} ^{(n)} = \mathcal{O} (\mathcal{E} ^{n})$.

To first order in the electric field, the Boltzmann equation becomes
\begin{equation}
(\partial _{t} + \tau _{\alpha \mathbf{k} } ^{-1}) \, \delta f _{\alpha} ^{(1)} = (e/ \hbar ) \, \mathbf{E} (t) \cdot \nabla _{\mathbf{k}} f _{0}
\end{equation}
Since the driving field oscillates harmonically at frequency $\omega$, the
first-order correction is naturally written as
\begin{equation}
\delta f _{\alpha} ^{(1)} (\mathbf{k} , t ) = h _{\alpha} (\mathbf{k}) \, e ^{i \omega t} + h _{\alpha} ^{\ast} (\mathbf{k}) \, e ^{-i \omega t} .
\end{equation}
Substitution into the Boltzmann equation immediately yields
\begin{align}
    h _{\alpha} (\mathbf{k}) = \frac{e \tau _{\alpha \mathbf{k} }}{2} \, \frac{\boldsymbol{\mathcal{E}} \cdot \mathbf{v} _{\alpha \mathbf{k}} }{1 + i \omega \tau _{\alpha \mathbf{k} } } \, f' _{0} (E _{\alpha \mathbf{k}})   , 
\end{align}
where $\mathbf v_{\alpha\mathbf{k}} = \frac{1}{\hbar} \nabla _{\mathbf k} E _{\alpha\mathbf{k}}$ is the band velocity.

The second-order correction satisfies
\begin{align}
    \left( \partial _{t} + \tau _{\alpha \mathbf{k} } ^{-1} \right) \, \delta f _{\alpha} ^{(2)} = \frac{e}{\hbar} \, \mathbf{E} (t) \cdot \nabla _{\mathbf{k}} \delta f _{\alpha} ^{(1)} . 
\end{align}
The source term contains two independent frequency components: a time-independent (rectified) contribution and another oscillating at twice the driving frequency. Accordingly, we write
\begin{align}
    \delta f _{\alpha} ^{(2)} (\mathbf{k} , t ) = g _{\alpha} (\mathbf{k} ) +  y _{\alpha} (\mathbf{k} ) \, e ^{2i \omega t} + \mbox{c.c.} \label{second_order_expansion}
\end{align}
where $g_\alpha$ describes the dc response, while
$y_\alpha$ gives the second-harmonic contribution.
Solving the Boltzmann equation yields
\begin{align}
    g _{\alpha} (\mathbf{k} ) = \frac{e \tau _{\alpha \mathbf{k} }}{2 \hbar } \, \boldsymbol{\mathcal{E}} \cdot \nabla _{\mathbf{k}} h _{\alpha} ^{\ast} (\mathbf{k}) , \label{g_definition} ,
\end{align}
and
\begin{align}
    y _{\alpha} (\mathbf{k} ) = \frac{e \tau _{\alpha \mathbf{k} }}{2 \hbar } \, \frac{\boldsymbol{\mathcal{E}} \cdot \nabla _{\mathbf{k}} h _{\alpha} (\mathbf{k})}{1 + 2 i \omega \tau _{ \alpha \mathbf{k} } }  .  \label{y_definition}
\end{align}
The gradients appearing in these expressions can be evaluated explicitly by differentiating the first-order solution. One obtains
\begin{align}
\boldsymbol{\mathcal E} \cdot \nabla _{\mathbf{k}} h _{\alpha} &= \frac{e}{2} \Bigg[ \frac{(\boldsymbol{\mathcal E} \cdot \mathbf{v} _{\alpha  \mathbf{k}}) \, f ' _{0} (\mathcal{E} _{\alpha \mathbf{k}}) }{(1 + i \omega \tau _{ \alpha \mathbf{k}}) ^{2}} \big( \boldsymbol{\mathcal E} \cdot \nabla _{\mathbf{k}} \tau _{\alpha \mathbf{k}} \big) \nonumber \\ & \phantom{=} + \frac{ \tau _{\alpha \mathbf{k}}}{1 + i \omega \tau _{\alpha \mathbf{k}}}\, \boldsymbol{\mathcal E} \cdot \nabla _{\mathbf{k}} \! \Big[(\boldsymbol{\mathcal E} \cdot \mathbf{v} _{\alpha \mathbf{k}}) \, f ' _{0} (\mathcal{E} _{\alpha \mathbf{k}}) \Big] \Bigg] , \label{eq:Egrad_h}
\end{align}
while the expression for
$\boldsymbol{\mathcal E} \cdot\nabla_{\mathbf k}h_\alpha^{*}$ follows by replacing $ i \rightarrow - i$.

The second term contains the band-curvature contribution, $\partial_{i}\partial_{j}E_{\alpha k}$, together with derivatives of the equilibrium distribution. By contrast, the first term depends explicitly on the momentum derivative of the relaxation time, $\nabla _{\mathbf{k}} \tau _{\alpha \mathbf{k}}$, showing that microscopic scattering mechanisms can directly influence the nonlinear response even before specifying a particular band structure. This rigorous treatment of $\tau$ as a strictly momentum-dependent quantity has been recently employed to accurately evaluate the thermoelectric performance of graphene \cite{3r17-kfy7,canas2026charge}.

Having obtained the nonequilibrium distribution up to second order in the electric field, we now evaluate the corresponding electric current. Within the semiclassical description, the charge current density is given by
\begin{align}
    \mathbf{J} (t) = - e \sum _{\alpha} \int \dot{\mathbf{r}} _{\alpha} \, f _{\alpha} (\mathbf{k},t) \, \frac{d ^{3} \mathbf{k}}{(2 \pi ) ^{3}} .
\end{align}
The first-order current reproduces the well-known linear conductivity and will not be discussed further here. Instead, we focus on the second-order response responsible for nonlinear Hall phenomena. At this order, the current naturally separates into two physically distinct contributions,
\begin{align}
    \mathbf{J} (t) = \mathbf{J} _{ \mbox{\scriptsize qp} } (t) + \mathbf{J} _{ \mbox{\scriptsize anom} } (t) , \label{Second_Order_Current}
\end{align}
where the first term originates from the group velocity weighted by the second-order correction to the distribution function,
\begin{align}
    \mathbf{J} _{ \mbox{\scriptsize qp} } (t) &= - e \sum _{\alpha} \int   \mathbf{v} _{\alpha \mathbf{k}} \; \delta f _{\alpha} ^{(2)} (\mathbf{k} , t )  \;  \frac{d ^{3} \mathbf{k}}{(2 \pi ) ^{3}} .  \label{Second_Order_Band}
\end{align}
while the second arises from the anomalous velocity associated with the Berry curvature acting on the first-order nonequilibrium distribution \cite{Sodemann_PRL2015},
\begin{align}
    \mathbf{J} _{ \mbox{\scriptsize anom} } (t) &= - \frac{e ^{2}}{\hbar} \, \sum _{\alpha} \int \mathbf{E} (t) \times \boldsymbol{\Omega} _{\alpha \mathbf{k}} \; \delta f _{\alpha} ^{(1)} (\mathbf{k} , t )  \; \frac{d ^{3} \mathbf{k}}{(2 \pi ) ^{3}} . \label{Second_Order_Anomalous}
\end{align}
These two mechanisms have different microscopic origins. The quasiparticle contribution reflects the modification of the carrier distribution by the external field, whereas the anomalous contribution originates entirely from the geometrical properties of the Bloch states encoded in the Berry curvature.

Because the applied field oscillates harmonically, the second-order current contains two independent frequency components. The first is a rectified (dc) current,
\begin{equation}
J_i^{(0)}
=
\sigma_{ijk}(0,\omega,-\omega)
E_jE_k^{*}
+\mathrm{c.c.},
\end{equation}
while the second oscillates at twice the driving frequency,
\begin{equation}
J_i^{(2\omega)}
=
\chi_{ijk}(2\omega,\omega,\omega)
E_jE_k
e^{2i\omega t}
+\mathrm{c.c.}, \label{second_hermonic_current}
\end{equation}
where
$\sigma_{ijk}$ and $\chi_{ijk}$ denote the rectification and
second-harmonic conductivity tensors, respectively.

In complete analogy with the current decomposition, each conductivity
tensor can be separated into quasiparticle and anomalous contributions,
\begin{align}
\sigma_{ijk}
&=
\sigma^{\rm qp}_{ijk}
+
\sigma^{\rm anom}_{ijk},
\\
\chi_{ijk}
&=
\chi^{\rm qp}_{ijk}
+
\chi^{\rm anom}_{ijk}.
\end{align}
After substituting the nonequilibrium distribution functions obtained in
the previous subsection and performing straightforward algebra, the
quasiparticle contribution to the rectified response becomes
\begin{equation}
\sigma^{\rm qp}_{ijk}
=
-
\frac{e^3}{4\hbar}
\sum_\alpha
\int
v_i
\tau_{\alpha\mathbf k}
\,
\partial_j
\left[
\frac{
v_k
\tau_{\alpha\mathbf k}
}
{
1-i\omega\tau_{\alpha\mathbf k}
}
f_0'
\right] \, \frac{d ^{3} \mathbf{k}}{(2 \pi ) ^{3}}  ,
\end{equation}
while the second-harmonic tensor reads
\begin{equation}
\chi^{\rm qp}_{ijk}
=
-
\frac{e^3}{4\hbar}
\sum_\alpha
\int
\frac{
v_i
\tau_{\alpha\mathbf k}
}
{
1+2i\omega\tau_{\alpha\mathbf k}
}
\,
\partial_j
\left[
\frac{
v_k
\tau_{\alpha\mathbf k}
}
{
1+i\omega\tau_{\alpha\mathbf k}
}
f_0'
\right] \, \frac{d ^{3} \mathbf{k}}{(2 \pi ) ^{3}}  .
\end{equation}
The anomalous contribution takes a particularly simple form. Remarkably, the same tensor governs both the rectified anomalous response and the second-harmonic anomalous conductivity \cite{Sodemann_PRL2015},
\begin{equation}
\xi_{ijk}(\omega)
\equiv
\sigma^{\rm anom}_{ijk}
=
\chi^{\rm anom}_{ijk},
\end{equation}
with
\begin{equation}
\xi_{ijk}(\omega)
=
\frac{e^3}{4\hbar}
\epsilon_{ilm}
\sum_\alpha
\int
\frac{
v_j
\tau_{\alpha\mathbf k}
\Omega_m
}
{
1+i\omega\tau_{\alpha\mathbf k}
}
f_0'(E_{\alpha\mathbf k}) \, \frac{d ^{3} \mathbf{k}}{(2 \pi ) ^{3}}   . \label{anomalous_tensor}
\end{equation}
Equations above constitute the central result of the semiclassical
formalism developed in this work. They show that the nonlinear Hall
response is completely determined by two ingredients: the geometric
properties of the electronic structure, encoded in the band velocity and
Berry curvature, and the transport relaxation time
$\tau_{\alpha\mathbf k}$, which contains all the microscopic information
about impurity scattering. In the following section we evaluate this
quantity for different impurity mechanisms, providing the microscopic
input required to determine the nonlinear Hall response.

\section{Relaxation times for different impurity mechanisms} \label{sec_relaxation_times}

The general conductivity tensors derived in the previous section show that the microscopic effects of disorder enter the nonlinear Hall response exclusively through the transport relaxation time $\tau _{\alpha \mathbf{k}}$. The purpose of this section is therefore to evaluate this quantity for different impurity mechanisms relevant to Weyl semimetals. We begin by introducing the low-energy model describing the electronic states around a Weyl node and subsequently derive the transport relaxation time for scalar and magnetic disorder within the Born approximation and beyond.

\subsection{Low-energy model of Weyl fermions}

The low-energy electronic structure of a Weyl semimetal is described by a pair of nondegenerate Weyl cones carrying opposite chiralities. Around each node, the quasiparticles are governed by the effective Hamiltonian
\begin{align}
H_{\chi}(\mathbf{k}) = \chi \hbar v_{F}  \boldsymbol{\sigma} \cdot \mathbf{k} + b _{0 \chi}, \label{Weyl_Hamiltonian}
\end{align}
where $v_{F}$ is the Fermi velocity, $\mathbf{k}$ is the crystal momentum measured from the node, $\boldsymbol{\sigma} = (\sigma_{x}, \sigma_{y}, \sigma_{z})$ are the Pauli matrices acting on a pseudospin (orbital or sublattice) degree of freedom, and $\chi = \pm 1$ denotes the chirality of the node \cite{RevModPhys.90.015001}.

The additional constant $b_{0\chi}$ accounts for a possible energy shift of each Weyl node. In the minimal two-node model, inversion symmetry constrains nodes of opposite chirality to lie at the same energy, whereas inversion breaking allows a relative energy displacement, $b_{0+}\neq b_{0-}$ \cite{Zyuzin_PRB2012}. In the present work, this energy displacement is introduced phenomenologically to generate different local Fermi energies at the two nodes.

The corresponding energy spectrum is
\begin{align}
    \mathcal{E} _{s \mathbf{k}} \equiv  b _{0 \chi} + s \mathcal{E} _{k} =  b _{0 \chi} + s \hbar v _{F} k
\end{align}
where $s = \pm 1$ denotes the conduction and valence bands. The associated normalized Bloch spinors are
\begin{align}
    \ket{\mathbf{k},\alpha } = \frac{1}{\sqrt{2}} \left( \begin{array}{c}
         \sqrt{1 + s \cos \theta }  \\[4pt]
          s \sqrt{1 - s \cos \theta } \, e ^{i\phi} 
    \end{array} \right)   , \label{Scatt_States} 
\end{align}
in spherical coordinates $\mathbf{k} = k \, ( \sin \theta \cos \phi , \sin \theta \sin \phi , \cos \theta)$. Using the Bloch states (\ref{Scatt_States}), it is straightforward to obtain the Berry curvature:
\begin{align}
    \boldsymbol{\Omega} _{\alpha \mathbf{k} }   = - s \chi \frac{\mathbf{k}}{2 k ^{3}} ,
\end{align}
indicating that each Weyl node acts as a monopole of Berry curvature in momentum space. These quantities completely determine the intrinsic geometric contribution to the nonlinear Hall response. The remaining ingredient is the transport relaxation time, which depends on the microscopic scattering mechanism and is evaluated below.

\subsection{General transport relaxation time}

Within the semiclassical Boltzmann approach, the transport relaxation time is determined by elastic scattering between Bloch states. For static disorder, the transition probability from an initial state $|\mathbf{k},\alpha\rangle$ to a final state $|\mathbf{k}',\alpha'\rangle$ is obtained from Fermi's golden rule,
\begin{align}
    \hspace{-0.2cm} W _{\mathbf{k} \alpha \to \mathbf{k}' \alpha '} = \frac{2 \pi }{\hbar }   \vert \langle \mathbf{k} ', \alpha ' \vert V (\mathbf{r}) \vert \mathbf{k} , \alpha \rangle \vert ^{2}  \delta (E _{ \alpha \mathbf{k}} - E _{ \alpha ' \mathbf{k}'}) , \label{Fermi_GR}
\end{align}
where $V( \mathbf{r})$ denotes the impurity potential.

The band- and chirality-resolved transport relaxation time is then defined as \cite{DasSarma_PRB2015,Ominato_PRB2014,Ominato_PRB2015}
\begin{align}
    \frac{1}{\tau _{ \alpha \mathbf{k}} } =  \int \frac{d ^{3} \mathbf{k}'}{(2 \pi ) ^{3}} \, W _{\mathbf{k} \alpha \to \mathbf{k}' \alpha } \, (1- \cos \theta _{\mathbf{k} \mathbf{k}'} ) , \label{relaxation_time}
\end{align}
with $\theta _{\mathbf{k} \mathbf{k}'} $ the scattering angle between incoming and outgoing momenta. The angular factor suppresses forward scattering contributions and ensures that $\tau _{ \alpha  \mathbf{k}}$ characterizes momentum relaxation, as opposed to the quasiparticle lifetime, which would follow from the same expression without the $(1- \cos \theta _{\mathbf{k} \mathbf{k}'} )$ term \cite{DasSarma_PRB2015}.

The explicit form of the relaxation time depends on the microscopic structure of the impurity potential. In the following subsections we consider successively scalar short-range, Gaussian, screened Coulomb, and magnetic impurities.

\subsection{Scalar disorder}

We first consider nonmagnetic impurities described by scalar potentials. Since the effective Hamiltonian in Eq. \eqref{Weyl_Hamiltonian} characterizes the low-energy excitations using the crystal momentum $\mathbf{k}$ measured directly from each Weyl node, it inherently operates within the envelope-function approximation. In this framework, the Weyl nodes are assumed to be well separated in momentum space, and the contact potentials introduced below act as effective coarse-grained scattering vertices rather than microscopic atomic point defects. Because these effective potentials lack the large momentum-transfer Fourier components required to couple distinct nodes \cite{Ominato_PRB2014}, they preserve the pseudospin degree of freedom and restrict the collision integral strictly to intranode processes. Such disorder constitutes the simplest microscopic mechanism for momentum relaxation and provides a useful baseline for the more general magnetic scattering processes discussed later.

The impurity potential is modeled as a collection of randomly distributed scattering centers,
\begin{align}
    V(\mathbf{r}) = \sum _{i=1} ^{N} u (\mathbf{r} - \mathbf{R} _{i}) , 
\end{align}
where $\mathbf{R} _{i}$ denote the impurity positions and $u (\mathbf{r})$ is the potential of a single scatterer. Different microscopic models are obtained by specifying the functional form of $u( \mathbf{r})$. For Weyl fermions described by the Bloch states (\ref{Scatt_States}), the spinor overlap is $\vert \langle \mathbf{k} ',\alpha '  \vert \mathbf{k} , \alpha  \rangle \vert ^{2} = \frac{1}{2} (1 + ss' \cos \theta _{\mathbf{k} \mathbf{k}'} ) $.

\subsubsection{Delta potential}

We first consider short-range impurities described by a contact interaction,
\begin{align}
    u (\mathbf{r}) = U _{0} \, \delta (\mathbf{r}),
\end{align}
where $U_0$ is the integrated contact strength and therefore has units of energy times volume. Its Fourier transform is
\begin{equation}
U (\mathbf{q}) = U _{0} .
\end{equation}
Substituting this potential into the general expression for the transport relaxation time yields
\begin{align}
    \frac{1}{\tau _{\alpha \mathbf{k}} ^{\delta}  } = \frac{  \pi n _{\mbox{\scriptsize i}} U _{0} ^{2} }{\hbar ^{2}  v _{F}}     \int \frac{d ^{3} \mathbf{k}'}{(2 \pi ) ^{3}} \, \sin ^{2} \theta _{\mathbf{k} \mathbf{k}'} \,  \delta  (k-k') ,  
\end{align}
which can be readily integrated in spherical coordinates,
leading to
\begin{align}
    \frac{1}{\tau _{\alpha \mathbf{k}} ^{\delta} } = \frac{ n _{\mbox{\scriptsize i}} U _{0} ^{2} \mathcal{E} _{k} ^{2}   }{3 \pi \hbar ^{4}  v _{F} ^{3} }   . \label{Rel_Time_Delta}
\end{align}
This result provides the characteristic quadratic energy
dependence expected for short-range disorder in Weyl
semimetals.

\subsubsection{Gaussian potential}

We next consider impurities with a finite spatial extent,
modeled by a Gaussian potential \cite{Ominato_PRB2014},
\begin{align}
    u (\mathbf{r}) = \frac{U _{0}}{(\sqrt{\pi} d ) ^{3}} \, e ^{- (r/d) ^{2}} , 
\end{align}
where $d$ characterizes the impurity size.
The corresponding Fourier transform is
\begin{equation}
U (\mathbf{q}) = U _{0} \, e ^{-(q/k _{0}) ^{2} } ,
\end{equation}
with $k_0=2/d$.

Introducing the characteristic energy
$\mathcal{E} _{0} = \hbar v _{F} k _{0}$, the transport relaxation time becomes
\begin{align}
    \frac{1}{\tau _{\alpha \mathbf{k}} ^{\mbox{\scriptsize G}} } &= \frac{  \pi n _{\mbox{\scriptsize i}} U _{0} ^{2} }{\hbar ^{2} v _{F} }  \int \frac{d ^{3} \mathbf{k}'}{(2 \pi ) ^{3}}   e ^{ - 8 ( \mathcal{E} _{k} / \mathcal{E} _{0} ) ^{2} \sin ^{2} (\theta _{\mathbf{k} \mathbf{k}'} / 2) } \notag \\ & \hspace{3cm} \times \sin ^{2} \theta _{\mathbf{k} \mathbf{k}'} \; \delta (k-k') ,
\end{align}
which can be written in the simple form
\begin{align}
    \tau _{\alpha \mathbf{k}} ^{\mbox{\scriptsize G}} = \tau _{\alpha \mathbf{k}} ^{\delta} \; \eta ( \mathcal{E} _{k} / \mathcal{E} _{0} )
\end{align}
where 
\begin{align}
    \eta ^{-1} (x) = \frac{3}{128 x ^{6} } \left[  ( 4x ^{2} - 1 ) + ( 4x ^{2} + 1 ) e ^{-8x ^{2}}  \right] . 
\end{align}
As expected, $\eta(x) \rightarrow 1$ when $\mathcal{E} _{k} / \mathcal{E} _{0} \rightarrow 0$,
recovering the contact-potential result in the limit
$d \rightarrow 0$.

\subsubsection{Screened Coulomb potential}

Charged impurities are naturally described by a screened Coulomb interaction \cite{Ominato_PRB2015,DasSarma_PRB2015}. In real space the corresponding Yukawa potential is
\begin{align}
    u (\mathbf{r}) = \frac{e ^{2}}{4 \pi \epsilon r} e ^{- q _{\mbox{\tiny TF}} r }
\end{align}
where $q _{\mbox{\scriptsize TF}}$ is the Thomas-Fermi screening wavevector and $\epsilon$ is the dielectric constant of the medium. The former is defined by $q _{\mbox{\scriptsize TF}} ^{2} = \frac{e ^{2}}{\epsilon} \nu (\mathcal{E} _{\alpha \mathbf{k}})$, where $\nu (\mathcal{E} _{\alpha \mathbf{k}})$ is the density of states. Its Fourier transform is
\begin{align}
    U(q) = \frac{e ^{2}}{\epsilon \, (q ^{2} + q _{\mbox{\tiny TF}}  ^{2} )} , 
\end{align}
which regularizes the long-wavelength divergence of the bare Coulomb interaction.

Substituting this potential into the transport equation gives
\begin{align}
    \frac{1}{\tau _{\alpha \mathbf{k}}  ^{\mbox{\scriptsize C}} } =    \frac{\pi ^{3}  v _{F} n _{\mbox{\scriptsize i}}  \alpha ^{2}}{ k ^{4} } \int \frac{d ^{3} \mathbf{k}'}{(2 \pi ) ^{3}} \, \frac{ \sin ^{2} \theta _{\mathbf{k} \mathbf{k}'} \, \delta (k-k')  }{\left[   \sin ^{2} (\theta _{\mathbf{k} \mathbf{k}'} / 2) + \frac{\alpha}{2 \pi} \right] ^{2}}   
\end{align}
where
\begin{align}
    \alpha
=
\frac{e^2}
{4\pi\varepsilon\hbar v_F} 
\end{align}
is the effective fine-structure constant.

After carrying out the angular integration one obtains
\begin{align}
    \tau _{\alpha \mathbf{k}}  ^{\mbox{\scriptsize C}}  = \frac{    \mathcal{E} _{k} ^{2}  }{4 \pi  ^{3}  \hbar ^{2} v _{F} ^{3} n _{\mbox{\scriptsize i}}  } \, \lambda ( \alpha / \pi )  ,  
\end{align}
with
\begin{align}
    \lambda (x) = \frac{1}{x ^{2}} \big[( 1 + x ) \, \mbox{arccoth} ( 1 + x ) - 1  \big] ^{-1} . 
\end{align}
The three scalar disorder models considered above differ quantitatively through the energy dependence of their relaxation times. Within the isotropic intranode model adopted here, however, they all produce a relaxation time that is independent of the Weyl-node chirality. Their contributions from opposite nodes can nevertheless fail to cancel when the nodes have different local Fermi energies. Magnetic impurities are considered next because they introduce a preferred direction and, beyond the first Born approximation, a helicity-dependent correction to the relaxation rate.

\subsection{Magnetic disorder}

For the magnetic-disorder analysis, we specialize to the ideal
spinful Weyl model in which the Pauli matrices
$\boldsymbol{\sigma}$ represent the physical-spin degree of freedom.
In this model, the Weyl Hamiltonian produces a three-dimensional
spin--momentum-locked texture, and localized magnetic moments couple
to the itinerant quasiparticles through the isotropic exchange
interaction in Eq.~\eqref{Magnetic_disorder}
\cite{Chang_PRB2015,Hosseini_PRB2015}. This minimal model is not
intended as a universal description of all Weyl semimetals, but it
provides a well-defined setting in which the interplay between the
Weyl spin texture and magnetic scattering can be investigated
analytically. If the Pauli matrices instead describe a purely orbital
or sublattice pseudospin without an associated real-spin texture, the
magnetic-scattering problem need not have the same structure, as also
emphasized in Ref.~\cite{Chang_PRB2015}.

Unlike scalar impurities, magnetic scatterers couple to the spinor texture of Weyl quasiparticles. Since the effective pseudospin and momentum are locked in a Weyl cone, this coupling introduces direction-dependent scattering and may distinguish states of opposite helicity beyond the first Born approximation. We show that unpolarized magnetic impurities produce identical transport relaxation times for both helicities. A finite polarization of the impurity moments introduces an anisotropic scattering channel; the helicity-dependent contribution arises from interference between the first- and second-order Born amplitudes, rather than from conventional skew scattering.

\subsubsection{First Born approximation}

We describe magnetic disorder by a collection of localized exchange
potentials,
\begin{align}
V (\mathbf{r})  = J \, \sum _{i=1} ^{N} \, \mathbf{S} _{i} \cdot \boldsymbol{\sigma} \; \delta(\mathbf{r}-\mathbf{R} _{i} ) , \label{Magnetic_disorder}
\end{align}
where $J$ is the exchange coupling, $\mathbf{S}_i$ is the localized magnetic moment of the $i$-th impurity, and $\mathbf{R}_i$ denotes its position. The local $s$--$d$ exchange interaction in Eq.~\eqref{Magnetic_disorder} is also the microscopic starting point for both RKKY and Kondo physics. In the RKKY mechanism, the itinerant Weyl quasiparticles mediate an effective interaction between distinct localized moments, which is generated at second order in $J$ after the electronic degrees of freedom are integrated out \cite{Chang_PRB2015,Hosseini_PRB2015}. By contrast, the Kondo problem concerns the many-body screening of an individual quantum magnetic impurity through repeated exchange scattering with the itinerant carriers \cite{Principi_PRB2015}. Neither effect is included in the present treatment: the impurity moments are regarded as prescribed and mutually independent scattering centers, and we calculate only the corresponding quasiparticle scattering amplitudes within a finite-order Born expansion.

The helicity basis diagonalizes the Weyl Hamiltonian and makes the spin-momentum locking explicit, providing a convenient framework to compute scattering amplitudes and to track chirality-dependent effects transparently. So in the following we work with the helicity spinors $\ket{\mathbf{k}, h }$ satisfying $(\boldsymbol{\sigma}\cdot\hat{\mathbf{k}})\ket{\mathbf{k}, h } = h \ket{\mathbf{k}, h } $ (with $h = s \chi$), which allow one to compute matrix elements analytically. The first-order Born approximation to the scattering amplitude is then given by
\begin{align}
\mathcal{M} ^{(1)}  _{\mathbf{k}'h', \mathbf{k}h}
= \bra{\mathbf{k} ', h '} V (\mathbf{r}) \ket{\mathbf{k}, h } , \label{M1_amplitude}
\end{align}
which can be evaluated analytically by inserting the explicit spinor wave functions. After straightforward algebra, one finds
\begin{align}
 \mathcal{M}^{(1)} _{\mathbf{k}'h', \mathbf{k}h} = \frac{J}{2   \mathcal{A} _{hh'} (\theta _{\mathbf{k}\mathbf{k}'})}   \tilde{\mathbf{S}} (\mathbf{q}) \cdot \left[ h   \hat{\mathbf{k}}    + h' \hat{\mathbf{k}}' +   i  h h' \hat{\mathbf{k}}' \times \hat{\mathbf{k}} \right]  , \label{Scatt_Magnetic_1order}
\end{align}
where $\mathcal{A} _{hh'} (\theta _{\mathbf{k}\mathbf{k}'}) \equiv \langle \mathbf{k} ', h ' \vert \mathbf{k}, h \rangle = \sqrt{( 1 +hh' \cos \theta _{\mathbf{k}\mathbf{k}'} ) / 2 }$ encodes the overlap between initial and final spinors, and $\tilde{\mathbf{S}} (\mathbf{q}) = \frac{1}{\mathcal{V}} \sum _{i=1} ^{N}   \mathbf{S} _{i}  e ^{i \mathbf{q} \cdot \mathbf{R} _{i} } $ is the spin density in momentum space, with $\mathbf{q} \equiv \mathbf{k}-\mathbf{k}'$ the momentum transfer. See the Appendix \ref{Appendix_Scatt_Amplitude_1er_order} for a detailed derivation of this expression. The amplitude resolves the helicity texture of the Weyl eigenstates. Nevertheless, after summing over the outgoing helicity and performing the impurity average, the first-Born transport rate remains independent of the incoming helicity, as shown explicitly below.

The first two terms in the scattering vertex are symmetric under
interchange of the incoming and outgoing states, whereas the amplitude
proportional to $\hat{\mathbf{k}}’\times\hat{\mathbf{k}}$ is
antisymmetric. Such a term produces conventional skew scattering only
if an antisymmetric contribution survives in the transition
probability, $W_{\mathbf{k}’\mathbf{k}}\neq
W_{\mathbf{k}\mathbf{k}’}$. Conventional skew scattering is
characterized by an antisymmetric scattering rate and generally arises
through interference between different orders of the scattering
amplitude \cite{Nagaosa_RMP2010}. As demonstrated below, no such
antisymmetric rate survives after the complete sum over outgoing
helicity in the present model.

For an ensemble of unpolarized magnetic impurities, specified by $\langle S _{i} \rangle = 0$ and $\langle S _{i} S _{j}\rangle = \frac{1}{3} S(S+1) \, \delta_{ij}$, the disorder-averaged squared matrix element takes the form
\begin{align}
     \langle \vert \mathcal{M}^{(1)} _{\mathbf{k}'h', \mathbf{k}h} \vert ^{2} \rangle  = \frac{n _{i} J ^{2} S(S+1)}{6} \left[ 2 + \frac{\sin ^{2} \theta _{\mathbf{k}\mathbf{k}'}  }{1 + h h' \cos \theta _{\mathbf{k}\mathbf{k}'} } \right] , \label{Average_Mag_Impurity}
\end{align}
as derived in Appendix \ref{Disorder_averaged_matrix_element}. Here, $n _{i}$ is the volume density of scatterers. To first-order Born approximation, the disorder-averaged differential rate depends on $h$ and $h'$ only through the product $hh'$. Summing Eq.~(\ref{Average_Mag_Impurity}) over the outgoing helicity
$h'=\pm1$, we obtain
\begin{align}
\sum_{h'=\pm1}
\langle \vert \mathcal{M}^{(1)} _{\mathbf{k}'h', \mathbf{k}h} \vert ^{2} \rangle
&=
n_iJ^{2}S(S+1).
\label{Average_Mag_Impurity_summed}
\end{align}
Thus, after summing over the outgoing helicity, the scattering probability is independent of both the incoming helicity and the scattering angle. Consequently, unpolarized magnetic impurities do not distinguish between the two incoming helicities at the level of the first Born approximation.

For an ensemble of polarized magnetic impurities with fixed moment $\tilde{\mathbf{S}} = S  \hat{\mathbf{m}}$ (no averaging over orientations), the first-order Born amplitude is given by Eq. (\ref{Scatt_Magnetic_1order}). Summing over the outgoing helicity one obtains the following contribution to the probability:
\begin{align}
     & \sum _{h'} \vert \mathcal{M}^{(1)} _{\mathbf{k}'h', \mathbf{k}h} \vert ^{2}  = \frac{ J ^{2} S ^{2}}{1 - \cos ^{2} \theta _{\mathbf{k}\mathbf{k}'} } \left[ (\hat{\mathbf{m}} \cdot \hat{\mathbf{k}} ) ^{2} + (\hat{\mathbf{m}} \cdot \hat{\mathbf{k}}' ) ^{2} \right. \notag \\ & \hspace{0.8cm} \left. \,  + (\hat{\mathbf{m}} \cdot \hat{\mathbf{k}} \times \hat{\mathbf{k}} ' ) ^{2} - 2 \cos \theta _{\mathbf{k}\mathbf{k}'} \, (\hat{\mathbf{m}} \cdot \hat{\mathbf{k}} ) (\hat{\mathbf{m}} \cdot \hat{\mathbf{k}}' ) \right] , \label{M1_unpolarized}
\end{align}
which is anisotropic with respect to the polarization axis $\hat{\mathbf{m}}$ but, crucially, contains no dependence on the sign of the incoming helicity $h$. Technical details of the derivation of Eq. (\ref{M1_unpolarized}) are presented in the Appendix \ref{app:polarized_magnetic_impurities}.  After integrating over final directions $\hat{\mathbf{k}}'$, and including the on-shell phase-space factor, the total (and transport) relaxation times remain identical for the two helicities.

\subsubsection{Beyond the Born approximation}

The equality of the two transport lifetimes at first order raises the possibility that multiple-scattering processes may generate a chirality-dependent correction. To examine this possibility, we employ the $T$-matrix expansion
\begin{align}
    T = V + V \, G(E) \, V +  V \, G(E) \, V \, G(E) \, V + \cdots , \label{T_matrix_expansion}
\end{align}
where $G(E)$ is the local Green function of the clean Weyl Hamiltonian. For a node of chirality $\chi$,
\begin{align}
    G _{\chi} (E) = \int \frac{d ^{3} \mathbf{p} }{(2 \pi ) ^{3}} \, \frac{E + \chi \hbar v _{F} \boldsymbol{\sigma} \cdot \mathbf{p} }{ E ^{2} - (\hbar v _{F} p) ^{2} + i0 ^{+} } . 
\end{align}
Rotational symmetry makes the vector part vanish after angular integration. The local propagator is therefore proportional to the identity matrix,
\begin{align}
    G _{\chi} (E) = G (E) \; \sigma _{0} ,  \label{Green}
\end{align}
with
\begin{equation}
G(E) = C _{\Lambda} (E) - i \pi \, \nu (E) \, \mbox{sgn}(E) .
\end{equation}
Here $\nu (E)$ is the density of states per Weyl node and
$C _{\Lambda} (E)$ denotes the cutoff-dependent principal-value part,
\begin{align}
    C _{\Lambda} (E) &=   \frac{E}{2 \pi ^{2} (\hbar v _{F} ) ^{3} } \left[ - \Lambda + \frac{E}{2} \, \ln \bigg| \frac{\Lambda + E}{\Lambda - E} \bigg| \, \right]  ,  
\end{align}
which provides a renormalization of scattering amplitudes that is blind to chirality, while the imaginary part is universal and encodes the phase acquired by on-shell intermediate states.

The second-order amplitude is
\begin{align}
\mathcal{M} ^{(2)}  _{\mathbf{k}'h', \mathbf{k}h}
= \bra{\mathbf{k} ', h '} V (\mathbf{r}) \, G (E) \, V (\mathbf{r}) \ket{\mathbf{k}, h } . \label{M2_amplitude}
\end{align}
For a local exchange potential, the Pauli-matrix identity $ ( \tilde{\mathbf{S}} \cdot \boldsymbol{\sigma} ) ^{2} = \tilde{\mathbf{S}} \cdot \tilde{\mathbf{S}} \, \sigma _{0} \equiv \tilde{S} ^{2} \, \sigma _{0}$ reduces the second-order amplitude to
\begin{align}
    \mathcal{M} ^{(2)}  _{\mathbf{k}'h', \mathbf{k}h}
= J ^{2} \, G(E) \, \tilde{S} ^{2} (\mathbf{q}) \, \mathcal{A} _{hh'} (\theta _{\mathbf{k}\mathbf{k}'})  . \label{Scatt_Magnetic_2order}
\end{align}
Up to this order, the scattering probability is
\begin{align}
    \vert \mathcal{M} _{\mathbf{k}'h', \mathbf{k}h} \vert ^{2} &= \vert \mathcal{M} ^{(1)}  _{\mathbf{k}'h', \mathbf{k}h} \vert ^{2} + \vert \mathcal{M} ^{(2)}  _{\mathbf{k}'h', \mathbf{k}h} \vert ^{2} \notag \\[5pt] & \hspace{1.5cm} + 2 \, \mbox{Re} \left[ \mathcal{M} ^{(1)}  _{\mathbf{k}'h', \mathbf{k}h} \, \mathcal{M} ^{(2) \ast} _{\mathbf{k}'h', \mathbf{k}h} \right] . \label{second_order_Born_app}
\end{align}
The first order contribution, when summed over the outgoing helicity, is given by the Eqs. (\ref{Average_Mag_Impurity_summed}) and (\ref{M1_unpolarized}) for unpolarized and polarized magnetic impurities, respectively. The second order contribution can be directly evaluated for unpolarized magnetic impurities. Using that $\big\langle |\tilde S^{2}(\mathbf q)|^{2} \big\rangle = n_i S^{4}$, 
the summation over the outgoing helicity yields  
\begin{align}
\sum _{h' } \langle \vert \mathcal{M}^{(2)} _{\mathbf{k}'h', \mathbf{k}h} \vert ^{2} \rangle
= J^{4}\,|G(E)|^{2}\,n_i S^{4},
\end{align}
which is manifestly independent of the initial helicity. Therefore, this term only renormalizes the amplitude without introducing chirality dependence. Moreover, the interference term between first- and second-order contributions vanishes upon disorder averaging. Explicitly, 
\begin{align}
\Big\langle  \mbox{Re} \left[ \mathcal{M} ^{(1)}  _{\mathbf{k}'h', \mathbf{k}h} \, \mathcal{M} ^{(2) \ast} _{\mathbf{k}'h', \mathbf{k}h} \right] \Big\rangle  \propto \langle  \tilde{\mathbf S}(\mathbf q) \, \tilde S ^{2 \ast }(\mathbf q)  \rangle , 
\end{align}
which is cubic in the spin variables. For an isotropic paramagnetic ensemble, all odd moments of the local spin vanish by symmetry. Consequently, both the pure second-order term and the averaged interference correction remain helicity blind for unpolarized magnetic impurities, and the two chiralities acquire identical relaxation times at this order.

We finally consider an ensemble with a finite polarization of the localized moments, $\tilde{\mathbf S}=S\hat{\mathbf m}$, where $\hat{\mathbf m}$ defines a preferred direction. In this case the orientational average is absent and the scattering probability is no longer rotationally invariant. At first order, summing the squared amplitude over the outgoing helicity gives Eq. (\ref{M1_unpolarized}). The pure second-order contribution also remains helicity independent. The technical details are relegated to the Appendix \ref{Appendix_M2_polarized}. For a fixed polarization,
\begin{align}
    \mathcal{M}^{(2)} _{\mathbf{k}'h', \mathbf{k}h} = J ^{2} S ^{2} G(E) A _{hh'}(\theta_{\mathbf{k} \mathbf{k} '}) , \label{M2}
\end{align}
and completeness of the outgoing helicity states gives
\begin{equation}
\sum _{h'} \left| A _{hh'}(\theta_{\mathbf{k} \mathbf{k} '}) \right| ^{2} = 1 .
\end{equation}
Consequently,
\begin{align}
\sum _{h' }  \vert \mathcal{M}^{(2)} _{\mathbf{k}'h', \mathbf{k}h} \vert ^{2} 
= J^{4}\,|G(E)|^{2}\,n_i S^{4}, \label{M2_summed}
\end{align}
which is manifestly independent of both the helicity $h$ and the polarization axis $\hat{\mathbf{m}}$. This contribution simply renormalizes the overall scattering rate without lifting the degeneracy between the two chiralities.

The interference between the first- and second-order Born amplitudes requires a more careful analysis than in the unpolarized case. Rather than manipulating explicitly the phases of the helicity spinors, it is convenient to evaluate the sum over the outgoing helicity by using the completeness relation of the Weyl eigenstates. As shown in Appendix~\ref{app:polarized}, this procedure leads to the remarkably simple result
\begin{equation}
\sum _{h'} \mbox{Re} \left[ \mathcal{M} ^{(1)}  _{\mathbf{k}'h', \mathbf{k}h} \, \mathcal{M} ^{(2) \ast} _{\mathbf{k}'h', \mathbf{k}h} \right]
=
hJ^3S^3
C _{\Lambda}(E) \, (\hat{\mathbf m}\cdot\hat{\mathbf k}),
\label{eq:interference_result}
\end{equation}
which depends only on the projection of the incoming momentum onto the polarization axis.

Equation~(\ref{eq:interference_result}) shows that the interference term does not generate a skew-scattering contribution proportional to $\hat{\mathbf m} \cdot (\hat{\mathbf k}'\times\hat{\mathbf k})$. Instead, it produces a helicity-dependent but anisotropic correction to the symmetric scattering rate.

Combining this interference term with the symmetric first- and second-order scattering probabilities, the transport relaxation rate becomes
\begin{align}
    \frac{1}{\tau _{ h \mathbf{k}} } &= \frac{2\pi n_i J ^{2} S ^{2} }{\hbar} \nu (E) \Big[ 1 + J ^{2} S ^{2} |G(E)| ^{2} \notag \\ & \hspace{3.3cm}  +  2hJ S C _{\Lambda} (E) \, ( \hat{\mathbf{m}} \cdot \hat{\mathbf{k}} ) \Big] . \label{relax_time_disordered} 
\end{align}
This equation constitutes the central result of the present Section. It shows that polarized magnetic impurities do not simply renormalize the transport lifetime but generate an anisotropic relaxation time whose value depends on the projection of the incoming quasiparticle momentum onto the impurity polarization axis. Since $h=s\chi$, the two Weyl chiralities acquire different relaxation times within a given band, providing a microscopic mechanism for chirality-dependent momentum relaxation. This anisotropic relaxation time will be employed directly in the transport calculations developed in the following Section.

\section{Disorder effects on nonlinear Hall conductivities in Weyl semimetals}
\label{Nonlinear_Hall_WSM_Section}

The microscopic scattering mechanisms discussed in the previous section enter the nonlinear transport problem exclusively through the transport relaxation time. Consequently, once the relaxation time has been determined for a given impurity model, its effect on the nonlinear Hall response can be incorporated without further approximations within the semiclassical formalism.

In this section we evaluate the rectified (dc) and second-harmonic nonlinear Hall conductivities of Weyl semimetals for the different impurity mechanisms considered in Sec.~\ref{sec_relaxation_times}. We first derive a general expression valid for an arbitrary momentum-dependent relaxation time. This formulation naturally separates the purely geometric contribution, determined by the Berry curvature, from the microscopic information encoded in the scattering processes. It also provides the appropriate framework for treating both isotropic scalar disorder and the anisotropic relaxation time generated by polarized magnetic impurities.

\subsection{General formalism}

For an isotropic Weyl cone described by the low-energy Hamiltonian (\ref{Weyl_Hamiltonian}), the quasiparticle contribution to the nonlinear Hall response vanishes identically. The band velocity is purely radial,
\begin{align}
    \mathbf{v}_{\alpha\mathbf{k}}
=
\frac{1}{\hbar}
\nabla_{\mathbf{k}}
\mathcal{E}_{\alpha\mathbf{k}}
=
sv_F\hat{\mathbf{k}},
\end{align}
while the equilibrium distribution depends only on the quasiparticle energy. Consequently, the quasiparticle tensors involve angular averages of odd-rank tensors such as $k_i$, $k_ik_jk_k$, or $k_i(k^2\delta_{jk}-k_jk_k)$, all of which vanish by rotational symmetry. Therefore,
\begin{align}
    \sigma^{ijk}_{\rm qp}(0,\omega,-\omega) = 0 , \qquad \chi^{ijk} _{\rm qp}(2\omega,\omega,\omega) = 0 ,
\end{align}
provided that the relaxation time is isotropic in momentum space. An anisotropic relaxation time, such as that generated by polarized
magnetic impurities, invalidates this angular cancellation and may
produce finite quasiparticle contributions.

The nonlinear Hall response of an isotropic Weyl semimetal is therefore entirely governed by the anomalous contribution arising from the Berry curvature. Moreover, the same tensor controls both the rectified and the second-harmonic responses,
\begin{align}
    \xi ^{ijk}(\omega) \equiv \sigma ^{ijk}_{\rm anom}(0,\omega,-\omega) = \sigma^{ijk}_{\rm anom}(2\omega,\omega,\omega),
\end{align}
reflecting the common microscopic origin of both effects through the anomalous velocity combined with the nonequilibrium correction to the distribution function.

For a given chemical potential, only one band crosses the Fermi level, so that the band index is fixed by
$s=\mathrm{sgn}(\mu-b_{0\chi})$. Introducing the positive Fermi energy measured from each Weyl node $\mu_\chi = |\mu-b_{0\chi}|$, the anomalous tensor (\ref{anomalous_tensor}) becomes
\begin{align}
\xi^{ijk}(\omega)
=
-
\frac{e^3v_F}{8\hbar}
\sum_\chi
\chi
\int
\frac{
\epsilon_{itk}
\tau _{ h \mathbf{k}} 
k_jk_t
}{
k^4
\left(
1+i\omega \tau _{ h \mathbf{k}} 
\right)
}
\tilde f_\chi'
(\mathcal E_k)
\frac{d^3  \mathbf{k} }{(2\pi)^3},
\label{eq:general_tensor}
\end{align}
where the relaxation time has been written explicitly as $\tau _{ h \mathbf{k}} $ to emphasize that the following derivation does not assume isotropic scattering. Here, Here,
\begin{align}
    \tilde f_\chi(E)
=
\frac{1}
{
e^{(E-\mu_\chi)/(k_BT)}+1
},
\qquad
\mu_\chi
=
|\mu-b_{0\chi}|,
\end{align}
is the Fermi-Dirac distribution written in terms of the chemical
potential measured from the energy of the node with chirality $\chi$.

Introducing the spherical coordinates $\mathbf{k} = k (\sin \theta \cos \phi , \sin \theta \sin \phi , \cos \theta ) $, with $d ^{3} \mathbf{k} = k^{2} dk d \Omega $, being $d \Omega $ the differential solid angle, the radial and angular integrations separate naturally,
\begin{align}
\xi^{ijk}(\omega)
=
-
\frac{e^3v_F}{8\hbar(2\pi)^3}
\sum_\chi
\chi
\int_0^\infty
\tilde f_\chi'(\mathcal E_k)
\,
\Pi^{ijk}_\chi(k,\omega) \,
dk ,
\label{eq:general_tensor2}
\end{align}
where we have introduced the angular tensor
\begin{align}
\Pi^{ijk}_\chi(k,\omega)
=
\int
d\Omega _{\mathbf{k}}
\, \epsilon_{itk}
\hat k_j
\hat k_t
\,
\frac{
\tau _{ h \mathbf{k}} 
}{
1+i\omega \tau _{ h \mathbf{k}}  
}.
\label{eq:Pi_general}
\end{align}
Equation~(\ref{eq:Pi_general}) constitutes the central result of this subsection. It shows that the nonlinear Hall response is completely determined by the angular structure of the relaxation time. Different impurity mechanisms enter only through the function $\tau _{ h \mathbf{k}}  $, while the Berry-curvature contribution remains unchanged.

For scalar disorder, the relaxation time depends only on the quasiparticle energy, $\tau _{ h \mathbf{k}}   = \tau(E_k)$, so that rotational symmetry allows the angular integration to be carried out analytically,
\begin{align}
    \Pi ^{ijk} _{\chi} = \frac{4\pi}{3} \epsilon _{ijk} \frac{\tau(E)}{1+i\omega\tau(E)} . 
\end{align}
Consequently, the anomalous tensor becomes $\xi ^{ijk}  ( \omega ) = \epsilon _{ijk} \, \psi (\omega ) $, where
\begin{align}
\psi(\omega)
=
\frac{e^3}{48\pi^2\hbar^2}
\sum_\chi
\chi
\int_0^\infty
\frac{\tau(E)}
{1+i\omega\tau(E)}
\left(
-
\frac{\partial\tilde f_\chi}
{\partial E}
\right)
dE.
\label{eq:psi_general}
\end{align}
In contrast, polarized magnetic impurities generate the anisotropic relaxation time derived in Sec.~\ref{sec_relaxation_times}, $\tau _{ h \mathbf{k}}$, which cannot be extracted from the angular integral. Consequently, the isotropic reduction leading to Eq.~(\ref{eq:psi_general}) no longer applies, and the nonlinear Hall response is promoted from the scalar function $\psi ( \omega)$ to the full tensor $\Pi^{ijk}_\chi$. The scalar and magnetic impurity models are analyzed separately in the following subsections.

\subsection{Scalar disorder}
\label{sec:scalar_nonlinear_response}

We first consider the scalar impurity models introduced in Sec.~\ref{sec_relaxation_times}. For delta-like, Gaussian, and screened Coulomb disorder, the relaxation time is independent of the Weyl-node chirality and of the direction of the quasiparticle momentum. It can therefore be written as $\tau _{ h \mathbf{k}}   = \tau(E_k)$, where the specific energy dependence is determined by the corresponding impurity potential. Under these conditions, the anomalous nonlinear conductivity retains the isotropic pseudotensor structure $\xi ^{ijk}  ( \omega ) = \epsilon _{ijk} \, \psi (\omega ) $, with the scalar response function given by
Eq.~(\ref{eq:psi_general}). 

\begin{figure*}[ht]
    \centering
    \captionsetup[subfigure]{font=scriptsize}
    \begin{subfigure}{0.32\textwidth}
         \centering
         \includegraphics[width=\linewidth]{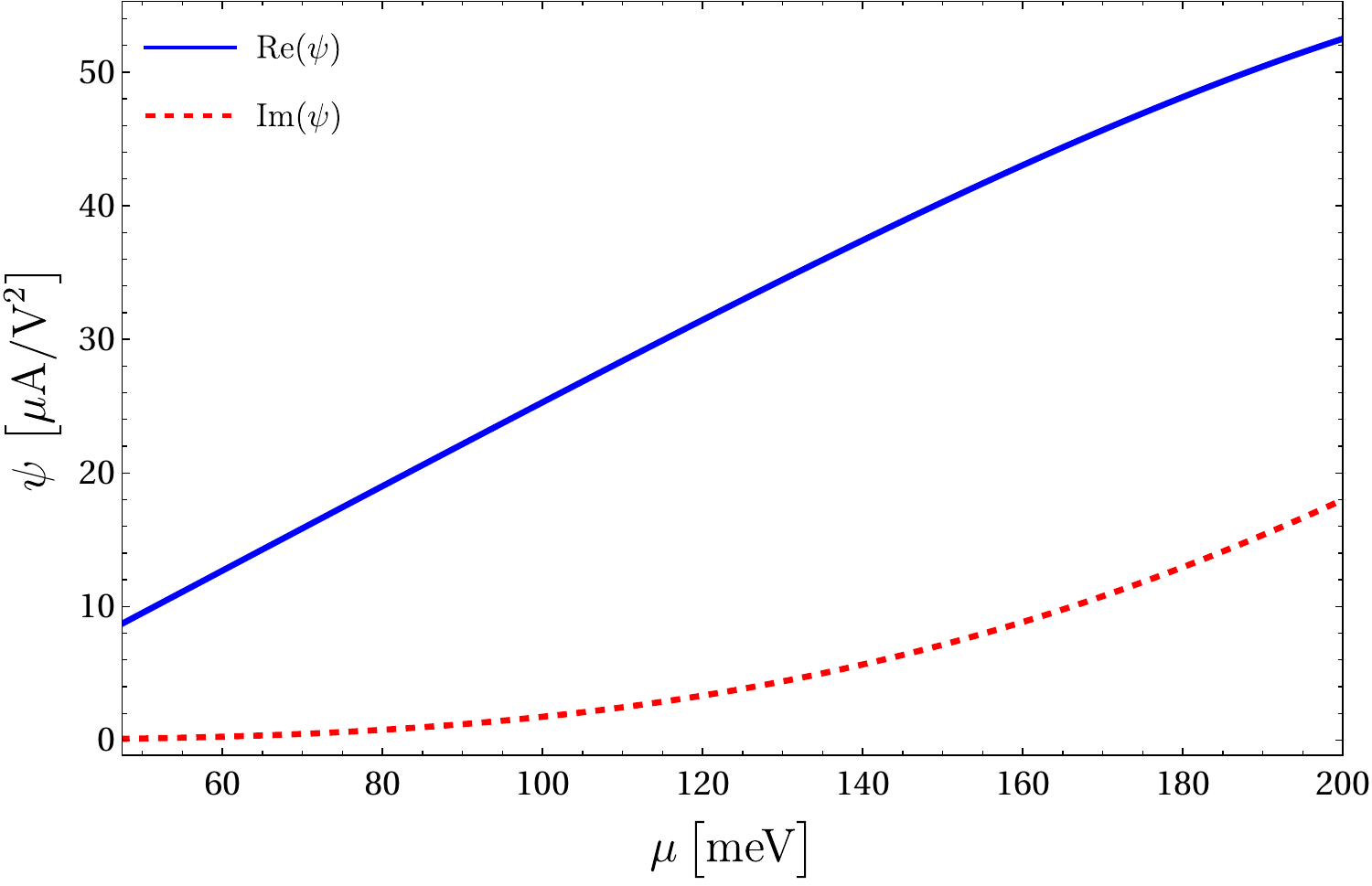}
         
         \subcaption{} \label{fig:Rect_Delta}
     \end{subfigure}\hspace{-.1 cm}
     \begin{subfigure}{0.32\textwidth}
         \centering
         \includegraphics[width=\linewidth]{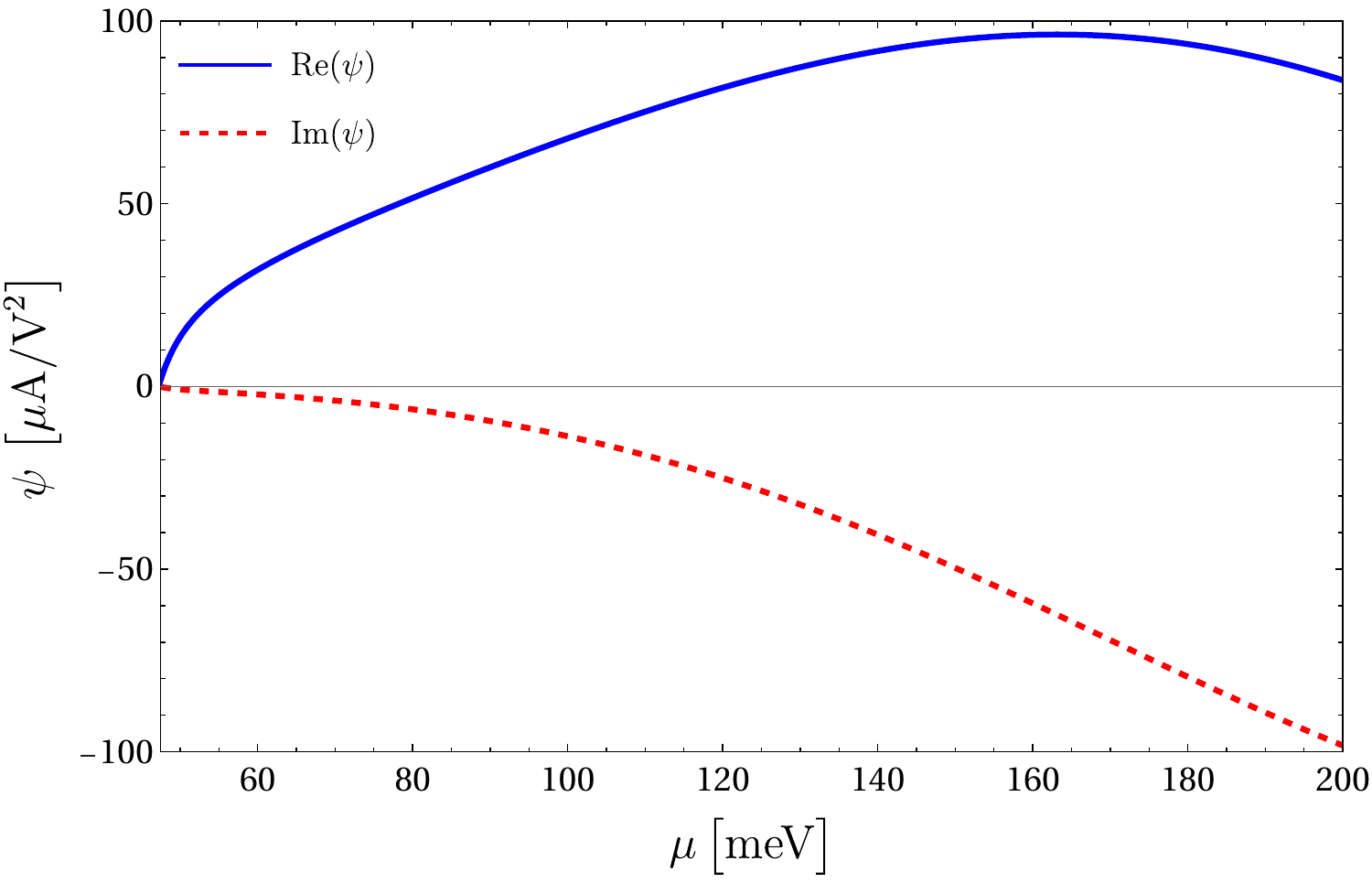}
         \subcaption{}\label{fig:Rect_Gauss}
     \end{subfigure}\hspace{-.1 cm}
     \begin{subfigure}{0.32\textwidth}
         \centering
         \includegraphics[width=\linewidth]{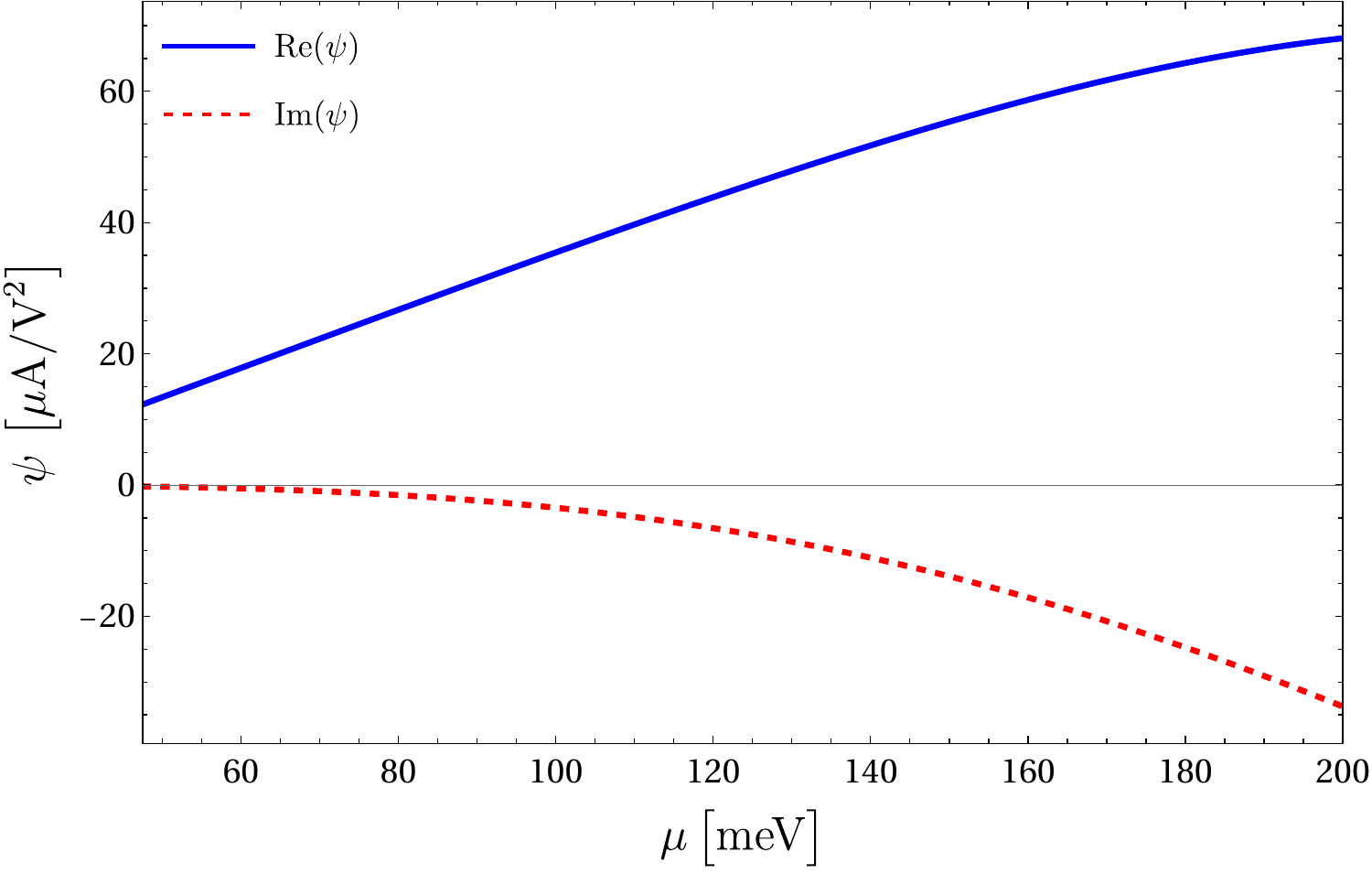}
         \subcaption{}\label{fig:Rect_Coul}
     \end{subfigure} 
     
     \begin{subfigure}{0.32\textwidth}
         \centering
         \includegraphics[width=\linewidth]{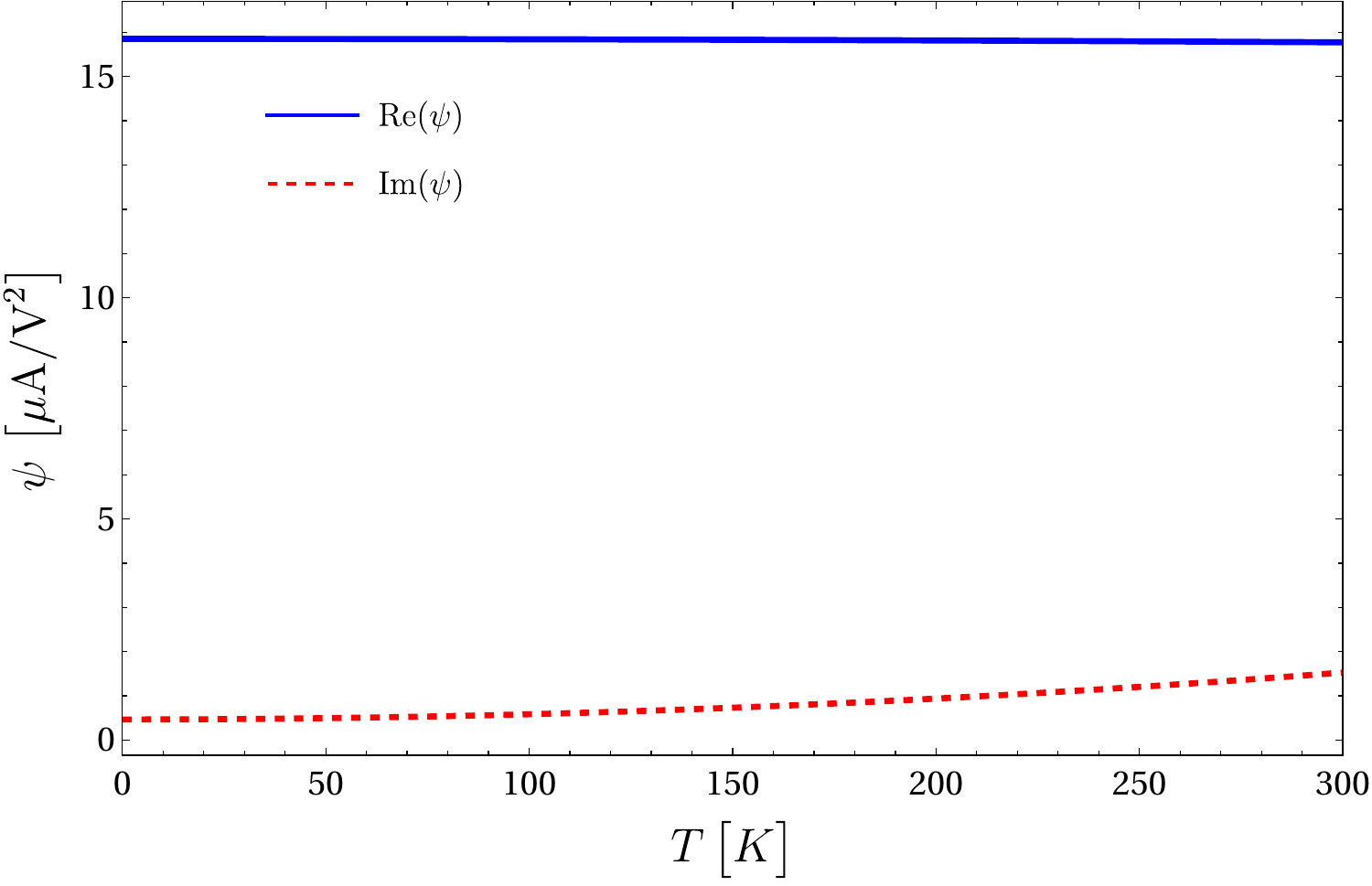}
         \subcaption{}\label{fig:RectT_Delta}
     \end{subfigure}\hspace{-.1 cm}
     \begin{subfigure}{0.32\textwidth}
         \centering
         \includegraphics[width=\linewidth]{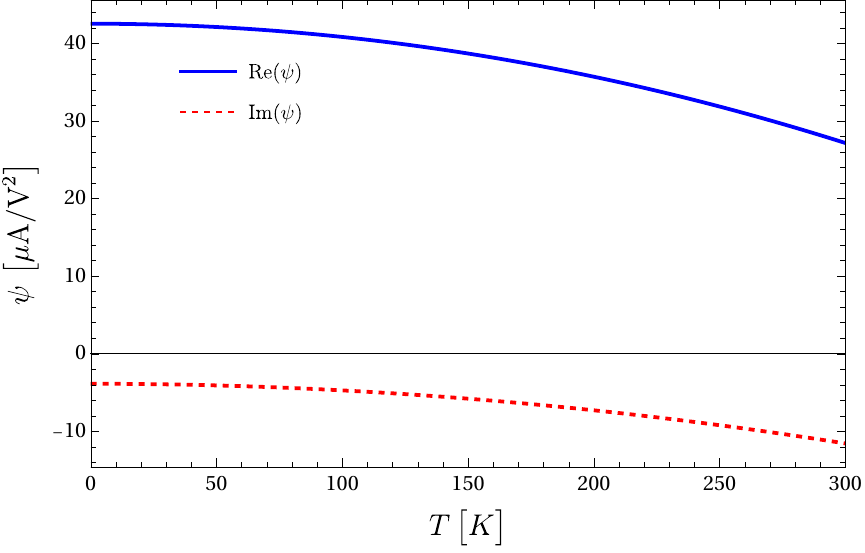}
         \subcaption{}\label{fig:RectT_Gauss}
     \end{subfigure}\hspace{-.1 cm}
     \begin{subfigure}{0.32\textwidth}
         \centering
         \includegraphics[width=\linewidth]{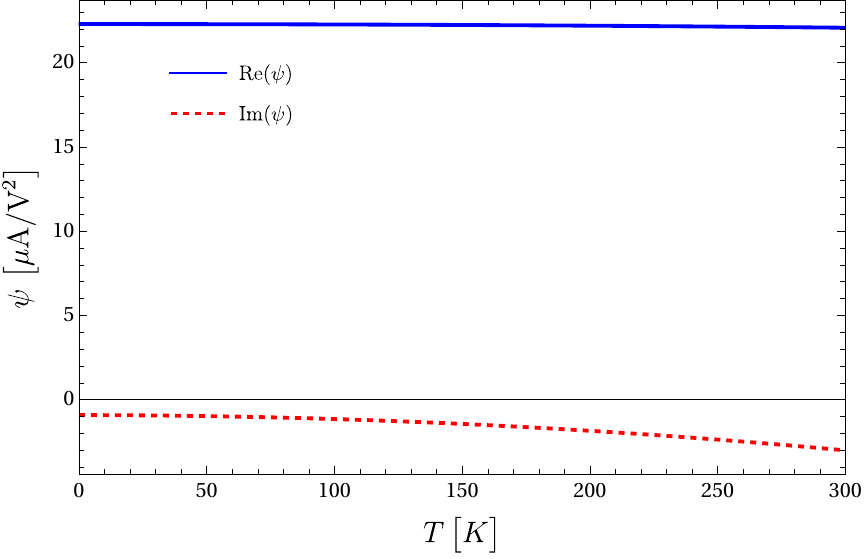}
         \subcaption{}\label{fig:RectT_Coul}
     \end{subfigure}
    \caption{Rectification efficiency $\psi$ for various impurity models. The nodal energy separation is set to $b_{0(+1)}-b_{0(-1)} = 40$ meV, and the Fermi energy $\mu$ is measured relative to the $\chi = -1$ node. Calculations are performed within the valid linear dispersion regime ($40 \leq \mu < 200$ meV), using $v_F = 10^5$ m/s, $n_\mathrm{i} = 10^{19}$ cm$^{-3}$, and the angular frequency
$\omega=10^{12}$ s$^{-1}$. The upper panels show the dependence of $\psi$ on $\mu$ at $T=0$ for (a) delta-function potentials ($U_0 = 30$ meV$ $\,nm$^3$), (b) Gaussian potentials ($d=4$ nm and $U_0$ so that $u(0)=30$ meV), and (c) screened Coulomb scatterers ($\epsilon = 3 \epsilon_0$). The lower panels (d, e, f) show the respective temperature dependence of $\psi$ at a fixed chemical potential $\mu = 70$ meV for the same impurity models.} 
 \label{fig:Rect_vs_mu}
\end{figure*}

Equation~(\ref{eq:psi_general}) shows that the two Weyl nodes contribute with opposite signs because their Berry curvatures carry opposite monopole charges. A finite response therefore requires that the two node contributions do not cancel exactly. In the present model, this occurs when the nodes lie at different energies, $b_{0+}\neq b_{0-}$, so that $\mu_+\neq\mu_-$ and the relaxation time is sampled at different Fermi energies for the two chiralities.

At zero temperature, the derivative of the Fermi-Dirac distribution, $-  \partial \tilde{f} _{\chi} (\mathcal{E} ) / \partial \mathcal{E} $, becomes strongly peaked around the Fermi energy $\mu _{\chi}$ and can be approximated by a Dirac delta function, $\delta (\mathcal{E} - \mu _{\chi} )$. This greatly simplifies the radial integral, since only states exactly at the Fermi level contribute to the nonlinear response. Explicitly, we can write
\begin{align}
    \psi  ( \omega , T = 0 ) =   \frac{e ^{3 } }{48 \hbar ^{2} \pi ^{2} }  \sum _{\chi} \chi \, \frac{ \tau ( \mu _{\chi} ) }{  1 + i \omega \tau (\mu _{\chi}) }   . 
\end{align}
This approximation highlights that, at zero temperature, the anomalous nonlinear Hall response is determined entirely by the properties of quasiparticles at the Fermi level, including the value of the momentum-dependent relaxation time $\tau (\mu _{\chi})$.

At low but finite temperatures, the radial integral can be evaluated using the Sommerfeld expansion, which provides a systematic approximation for integrals weighted by the derivative of the Fermi-Dirac distribution. For a smooth function $g(\mathcal{E})$, the expansion reads
\begin{align}
    \int _{0} ^{\infty} g(\mathcal{E}) \,  \left( -  \frac{\partial \tilde{f} _{\chi}  (\mathcal{E})}{\partial \mathcal{E} } \right) d \mathcal{E} & \approx g(\mu _{\chi}) + \frac{\pi ^{2}}{6} (k _{B}T) ^{2} g'' (\mu _{\chi}) \notag \\ & \hspace{1.3cm} + \mathcal{O} [ (k _{B}T) ^{4} ] ,
\end{align}
where $\mu _{\chi}$ is the chemical potential as measured from the node energy, $k _{B}$ is Boltzmann's constant, and $g'' (\mu _{\chi})$ is the second derivative of $g$ evaluated at $\mathcal{E} = \mu _{\chi}$. Applying this to the energy-dependent relaxation-time integral, we identify $g(\mathcal{E}) = \frac{ \tau (\mathcal{E} ) }{  1 + i \omega \tau (\mathcal{E} ) } $. Thus, the integral can be approximated as
\begin{align}
    \psi  ( \omega , T ) &=  \frac{e ^{3 } }{48 \hbar ^{2} \pi ^{2} } \sum _{\chi} \chi \;  \Bigg\{\ \frac{ \tau ( \mathcal{E} ) }{  1 + i \omega \tau ( \mathcal{E} ) }  +  \frac{ \pi ^{2} }{ 6 }  \,  (k _{B}T) ^{2}  \notag \\ & \hspace{1.5cm} \times \frac{d ^{2}}{d \mathcal{E} ^{2}} \left[  \frac{ \tau ( \mathcal{E} ) }{  1 + i \omega \tau (\mathcal{E}) } \right] \Bigg\}\ \Bigg| _{\mathcal{E} = \mu _{\chi}} . \label{psi_function} 
\end{align}
The first term reproduces the zero-temperature result, while the second term captures the leading finite-temperature correction. This formulation clearly shows that, at low temperatures, the nonlinear Hall response is dominated by states near the Fermi level, with temperature effects entering through the curvature of the energy dependence of the relaxation time.

With the previous results at hand, we are now in a position to evaluate the nonlinear currents. In particular, using the explicit form of the rectified (dc) and second-harmonic tensors, namely $\sigma^{ijk}_{\rm anom}(0,\omega,-\omega)=\sigma^{ijk}_{\rm anom}(2\omega,\omega,\omega)=\xi^{ijk}(\omega)=\epsilon_{ijk}\,\psi(\omega,T)$ with $\psi (\omega , T)$ given by Eq. (\ref{psi_function}), the corresponding charge currents can be written as follows. Importantly, the second-harmonic current [Eq.~(\ref{second_hermonic_current})], which takes the form $\mathbf{J} (2 \omega ) = 2 \, \mbox{Re} \, \left[ \psi (\omega ,T ) \, \boldsymbol{\mathcal{E}} \times \boldsymbol{\mathcal{E}} e ^{2i \omega t} \right]$ vanishes identically in the isotropic Weyl model. However, the rectified (dc) current remains finite and is given by
\begin{align}
    \mathbf{J} (0) = 2 \, \mbox{Re} \, \left[ \psi (\omega ,T ) \, \boldsymbol{\mathcal{E}} \times \boldsymbol{\mathcal{E}} ^{\ast} \right] . \label{dc_current}
\end{align}
In this expression, the function $\psi (\omega,T)$ encodes the microscopic details of the electronic structure and impurity scattering in the Weyl semimetal. It has the dimensions of conductivity divided by frequency, ensuring that the resulting current has the correct units of charge per unit time per unit area. Physically, $\psi (\omega,T)$ quantifies the efficiency of the rectification process under optical driving, with its dependence on $\omega$ and $T$ reflecting the combined effects of dynamical response and thermal smearing near the Fermi level. In SI units, since $\mathbf{J}$ has units A/m$^{2}$ and $\boldsymbol{\mathcal{E}} \times \boldsymbol{\mathcal{E}} ^{\ast}$ has units (V/m) $^{2}$, one has
\begin{align}
    \left[ \psi \right] = \frac{\mbox{A }}{\mbox{V} ^{2}} , 
\end{align}
which is consistent with the prefactor $e^{3}\tau / \hbar^{2}$ appearing in our expressions. These are the standard units for a second-order conductivity. Importantly, the structure of Eq.~\eqref{dc_current} shows that the rectified current is proportional to
$\boldsymbol{\mathcal E}\times\boldsymbol{\mathcal E}^{*}$, which
vanishes for linearly polarized light. For circularly polarized light,
this vector is finite and reverses sign with the optical helicity. This
is the characteristic field structure entering helicity-dependent
circular photogalvanic responses in Weyl semimetals
\cite{Golub_PRB2018}.

In Fig. \ref{fig:Rect_vs_mu} we plot the rectification efficiency, $\psi(\omega,T)$, as a function of the Fermi energy $\mu$ for different types of impurities, including delta-like, Gaussian and Coulomb potentials. Both the real and imaginary parts of $\psi(\omega,T)$ are shown, illustrating how the microscopic nature of disorder affects the amplitude and phase of the rectified current. As derived in Sec. III, the net rectification arises from the incomplete cancellation of the chiral nodes' contributions. Because the relaxation times $\tau(\mathcal{E})$ exhibit distinct functional dependencies on energy for each impurity model, the nodal energy separation ($b_{0+} \neq b_{0-}$) induces a strong asymmetry in the scattering rates at the respective Fermi levels, yielding $\tau(\mu_+) \neq \tau(\mu_-)$. 

It is important to emphasize that the physical parameters used in our calculations serve as a phenomenological order-of-magnitude guide to capture the qualitative transport physics across a broad family of Weyl semimetals, rather than modeling a single specific material. For instance, the nodal energy separation of $40$ meV is consistent with values reported for type-II Weyl semimetals like MoTe$_2$ \cite{DengNP2016}, whereas the validity range of the linear Dirac dispersion ($\mu < 200$ meV) is taken from prototypical systems like BiO$_2$ \cite{Young_PRL2012}. Similarly, the chosen Fermi velocity ($v_F = 10^5$ m/s) represents a typical magnitude for these topological materials \cite{RevModPhys.90.015001}. Furthermore, the selected impurity density ($n_\mathrm{i} = 10^{19}$ cm$^{-3}$) corresponds to a dilute doping regime ($\sim 0.1\%$) that ensures the validity of the Born approximation, yielding finite transport relaxation times without destroying the intrinsic topological band structure. The validity of the
weak-scattering Born--Boltzmann treatment must be checked a posteriori
through conditions such as $\Gamma(\mu_\chi)\ll\mu_\chi$ and
$k_{F,\chi}\ell_\chi\gg1$, rather than inferred from the impurity
concentration alone \cite{Ominato_PRB2014,Ominato_PRB2015}. Ultimately, the plots
should therefore be interpreted as illustrating generic
disorder-dependent trends within the assumed continuum and
weak-scattering regimes.

Importantly, the structure of Eq.~(\ref{dc_current}) shows that the rectified current is proportional to the vector $\boldsymbol{\mathcal{E}} \times \boldsymbol{\mathcal{E}}^{\ast}$, which identically vanishes for linearly polarized light. In contrast, for circularly polarized light this vector is finite and points along the propagation direction of the field, leading to a nonzero rectified response. This highlights that the nonlinear Hall effect in the isotropic Weyl model is intrinsically tied to the helicity of the optical field. To see this explicitly, let us consider a Weyl semimetal under the influence of a plane wave circularly polarized, i.e.
\begin{align}
    \boldsymbol{\mathcal{E}} = E _{0} \left( \begin{array}{c}
          \cos \theta  \\ i \lambda \\ \sin \theta
    \end{array} \right) , 
\end{align}
where $\lambda = \pm 1$ determines if the wave is right or left handed polarized. Inserting this expression now in Eq. (\ref{dc_current}) one gets
\begin{align}
    \mathbf{J} (0) =  4 \lambda \vert E _{0} \vert ^{2} \left( \begin{array}{c}
          - \sin \theta  \\ 0 \\ \cos \theta
    \end{array} \right) \, \mbox{Im} \left[ \psi (\omega ,T )  \right] .
\end{align}
In summary, within the present two-node model, scalar disorder produces
a net rectified response only when the contributions from opposite
chiralities are inequivalent, for example because the nodes lie at
different energies. Polarized magnetic impurities modify the angular
and tensorial structure of the scattering rate and generate a
helicity-dependent correction beyond the first Born approximation.
They do not, however, prevent cancellation between otherwise identical
and energetically equivalent Weyl nodes.

\subsection{Polarized magnetic impurities}
\label{sec:polarized_magnetic_response}

We now consider the nonlinear response generated in the presence of polarized magnetic impurities. In contrast to scalar disorder, the relaxation time depends not only on the quasiparticle energy but also on the helicity and on the propagation direction relative to the impurity polarization axis.

As obtained in Sec.~\ref{sec_relaxation_times}, the transport relaxation rate is given by Eq. (\ref{relax_time_disordered}), and the factor involving the relaxation time takes the particularly simple form
\begin{align}
\frac{\tau_{h\mathbf{k}}}{1+i\omega\tau_{h\mathbf{k}}} = \frac{1}{A(E_k,\omega) + h B(E _{k}) (\hat{\mathbf m}\cdot\hat{\mathbf k})}, \label{eq:magnetic_dynamic_factor}
\end{align}
where
\begin{align}
A(E,\omega ) &= \frac{2\pi n_iJ^2S^2}{\hbar} \nu(E) \left[ 1+J^2S^2|G(E)|^2 \right] + i \omega , \label{eq:A_magnetic} \\ B(E) &= \frac{4\pi n_iJ^3S^3}{\hbar} \nu(E) C _{\Lambda} (E) . \label{eq:B_magnetic}
\end{align}
Substituting Eq.~(\ref{eq:magnetic_dynamic_factor}) into the angular
tensor (\ref{eq:Pi_general}) gives
\begin{align}
\Pi ^{ijk} _{h} (k,\omega ) = \epsilon _{itk} \int d \Omega _{\mathbf{k}} \, \frac{   \hat{k} _{j} \hat{k} _{t} }{ A  + h B  \, (\hat{\mathbf{m}} \cdot \hat{\mathbf{k}} ) } , \label{eq:Pi_magnetic_integral}
\end{align}
where we have suppressed the dependence of $A$ and $B$ on $E_{k}$ for simplicity. Because the impurity polarization introduces a single preferred axis, the second-rank angular integral appearing in Eq.~(\ref{eq:Pi_magnetic_integral}) must have the axial form
\begin{align}
\int d \Omega _{\mathbf{k}} \, \frac{\hat{k} _{j} \hat{k} _{t} }{A+h B \, (\hat{\mathbf{m}} \cdot \hat{\mathbf{k}} ) } = \mathcal{P} _{\perp,h} \left( \delta_{jt} - \hat{m} _{j} \hat{m} _{t} \right) + \mathcal{P} _{\parallel,h} \hat{m} _{j} \hat{m} _{t} , \label{eq:axial_angular_decomposition}
\end{align}
where the transverse and longitudinal angular functions are
\begin{align}
\mathcal{P} _{\parallel,h}  = \int _{-1} ^{1} \frac{2 \pi  u ^{2} \, du}{A+h B  u}  , \quad \mathcal{P} _{\perp,h} = \pi \int _{-1} ^{1} \frac{(1-u ^{2}) \, du}{A+h B  u} ,
\end{align}
respectively. These angular integrals can be evaluated analytically, yielding
\begin{align}
\mathcal{P} _{\parallel,h} (E,\omega ) &= \frac{2 \pi hA}{B^3}   \left[ A \ln \left( \frac{A+B h}{A-B h} \right) -2 B h \right] , \label{eq:Pparallel_integral} \\[5pt] \mathcal{P} _{\perp,h} (E,\omega ) &=  \frac{ \pi A}{h B^3} \left[ 2 h B  - \frac{ (A ^{2} - B ^{2})  }{ A } \ln \left( \frac{A+B h}{A-B h} \right) \right] .
\end{align}

\begin{figure*}[ht]
    \centering
    \begin{subfigure}{0.32\textwidth}
         \centering
         \includegraphics[width=\linewidth]{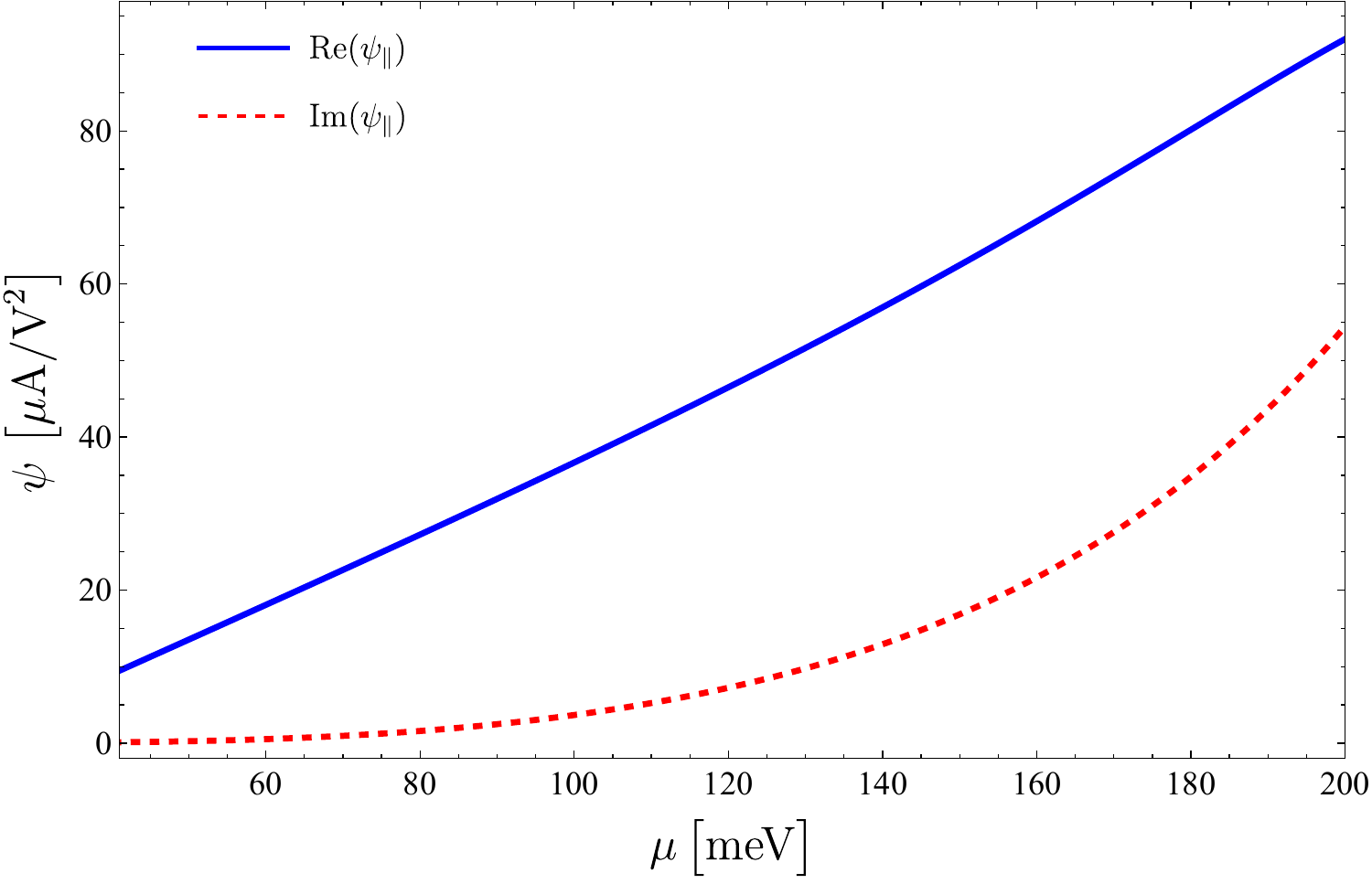}
         \subcaption{}\label{fig:Rect_Mag}
     \end{subfigure}
     \begin{subfigure}{0.32\textwidth}
         \centering
         \includegraphics[width=\linewidth]{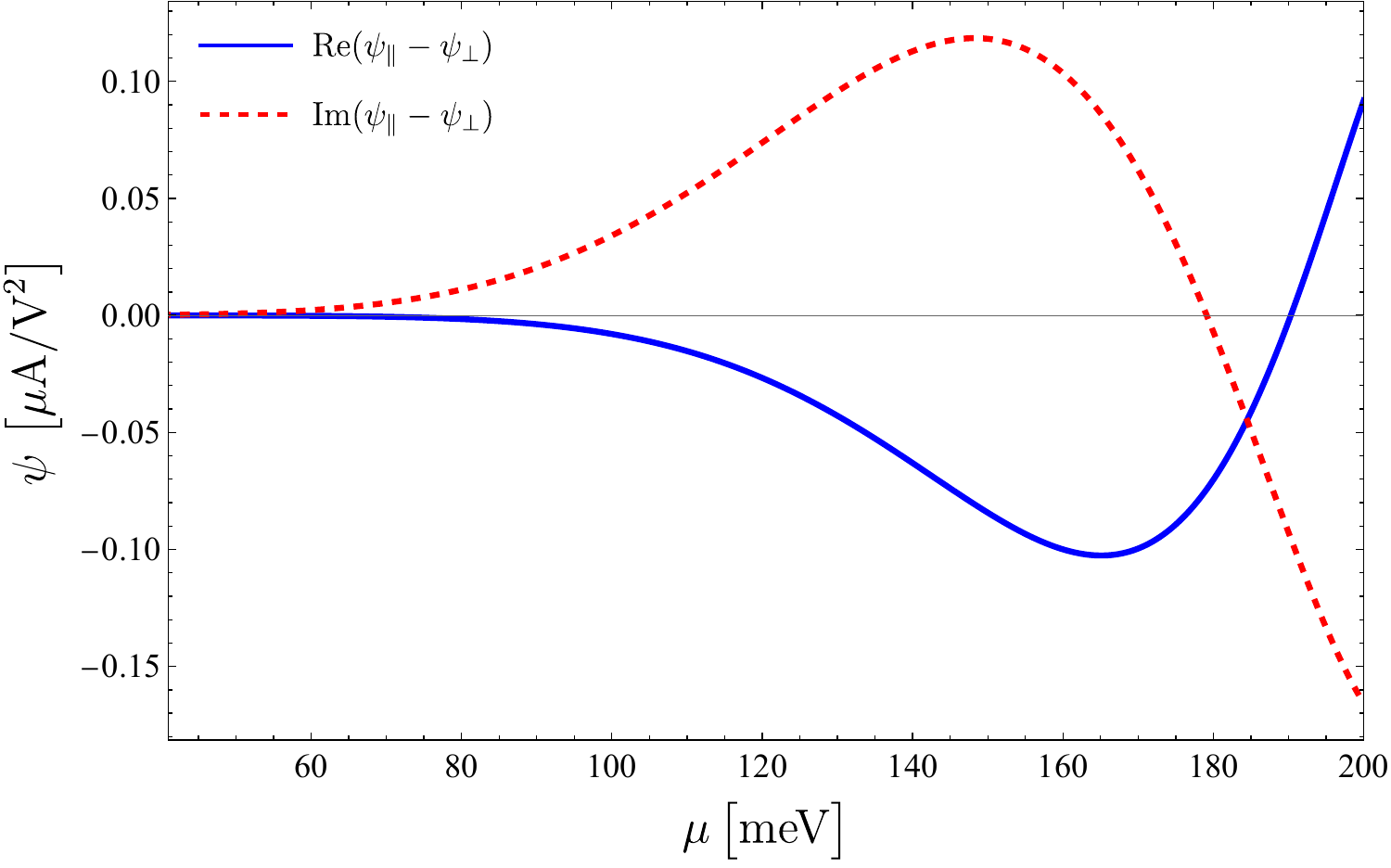}
         \subcaption{}\label{fig:Rect_Mag_diff}
     \end{subfigure}
     
     \begin{subfigure}{0.32\textwidth}
         \centering
         \includegraphics[width=\linewidth]{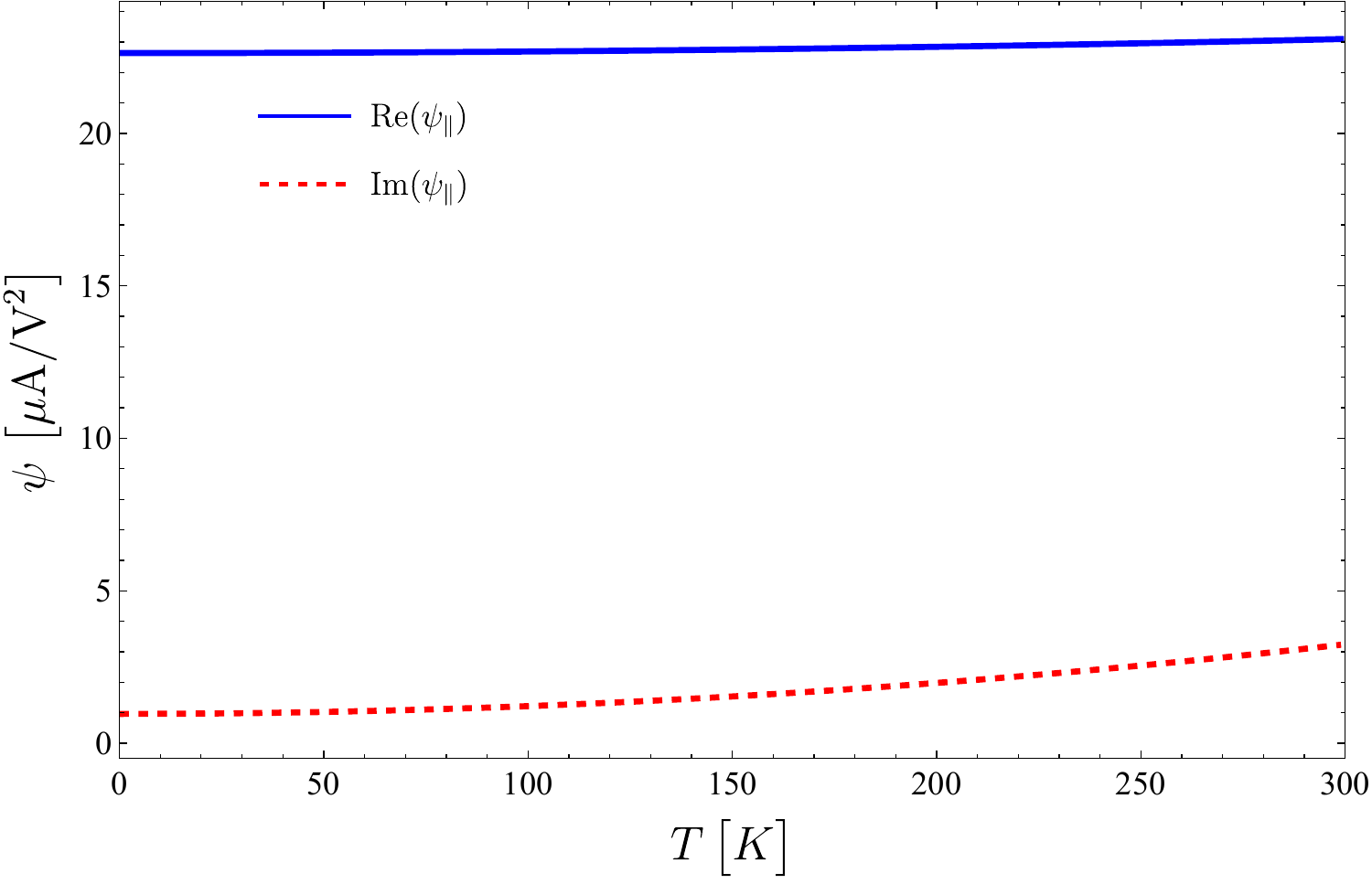}
         \subcaption{}\label{fig:RectT_Mag}
     \end{subfigure}
     \begin{subfigure}{0.32\textwidth}
         \centering
         \includegraphics[width=\linewidth]{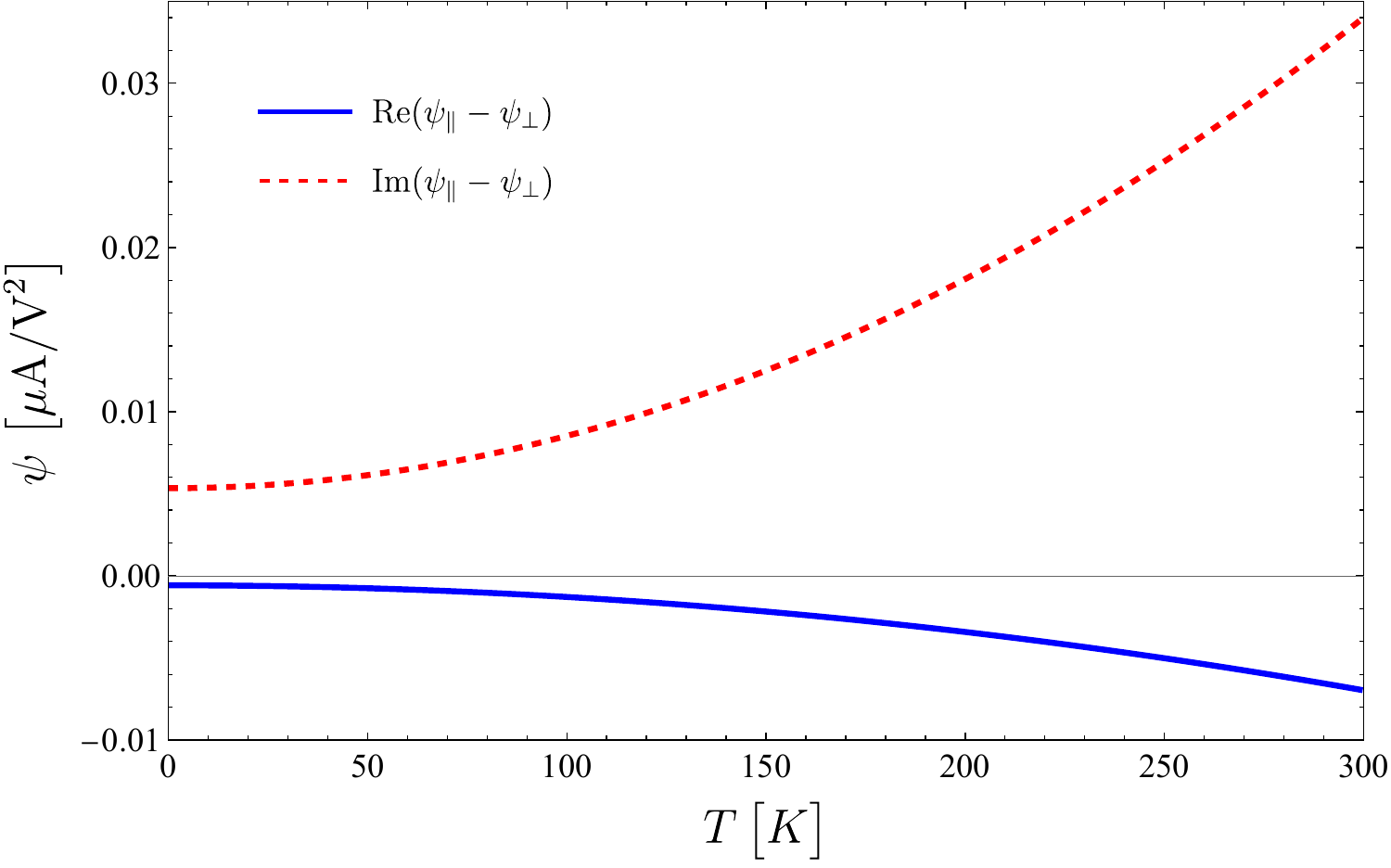}
         \subcaption{}\label{fig:RectT_Mag_diff}
     \end{subfigure}
     
    \caption{Rectification efficiency for polarized magnetic impurities. The figure is arranged in a $2 \times 2$ grid. The upper panels display the dependence on the Fermi energy $\mu$ at $T=0$ for (a) the longitudinal component $\psi_\parallel$ and (b) the anisotropic difference $\psi_\parallel - \psi_\perp$. The lower panels show the temperature dependence at a fixed chemical potential $\mu = 70$ meV for (c) $\psi_\parallel$ and (d) $\psi_\parallel - \psi_\perp$. As observed, the difference $\psi_\parallel - \psi_\perp$ is finite but 3 to 4 orders of magnitude smaller than $\psi_\parallel$; therefore, $\psi_\perp$ is practically identical to $\psi_\parallel$ and is not plotted separately. he continuum exchange strength is set to $(JS)_{\rm cont}=20.6$ meV\,nm$^3$, obtained by combining the exchange energy scale $\Delta_{\rm ex}=165$ meV used for EuCd$_2$As$_2$ \cite{Jo_NatC2021} with the primitive-cell volume $V_{\rm cell}\simeq0.125$ nm$^3$ \cite{Wang_PRB2019}. All other physical parameters are identical to those in Fig. \ref{fig:Rect_vs_mu}.} 
 \label{fig:RectM}
\end{figure*}

Using Eq.~(\ref{eq:axial_angular_decomposition}), the angular tensor
takes the form
\begin{align}
\Pi^{ijk}_\chi
=
\mathcal P_{\perp,\chi}\,
\epsilon_{ijk}
+
\left(
\mathcal P_{\parallel,\chi}
-
\mathcal P_{\perp,\chi}
\right)
\epsilon_{itk}m_jm_t.
\label{eq:Pi_magnetic_decomposition}
\end{align}
Therefore, the anomalous nonlinear conductivity can be expressed as
\begin{align}
\xi^{ijk}(\omega)
=
\epsilon_{ijk}\,
\psi_\perp(\omega)
+
\epsilon_{itk}m_jm_t
\left[
\psi_\parallel(\omega)
-
\psi_\perp(\omega)
\right],
\label{eq:xi_magnetic_decomposition}
\end{align}
where
\begin{align}
\psi_a(\omega,T)
=
\frac{e^3}
{8\hbar^2(2\pi)^3}
\sum_{\chi}
\chi
\int_0^\infty
\mathcal P_{a,h}(E,\omega)
\left(
-
\frac{\partial\tilde f_\chi }
{\partial E}
\right)
dE,
\label{eq:psi_magnetic_general}
\end{align}
where $a=\perp,\parallel$. Equation~(\ref{eq:xi_magnetic_decomposition}) shows that polarized magnetic impurities promote the isotropic scalar response $\psi(\omega)$ into two independent response functions: $\psi_\perp$, associated with directions transverse to the impurity polarization, and $\psi_\parallel$, associated with the longitudinal direction. The scalar-disorder result is recovered when $\psi_\parallel=\psi_\perp$.

At zero temperature, only the Fermi surfaces of the two Weyl nodes
contribute, and Eq.~(\ref{eq:psi_magnetic_general}) becomes
\begin{align}
\psi_a(\omega,0)
=
\frac{e^3}
{8\hbar^2(2\pi)^3}
\sum_{\chi }
\chi\,
\mathcal P_{a,h}(\mu_\chi,\omega) . 
\label{eq:psi_magnetic_zeroT}
\end{align}
At low but finite temperature, a Sommerfeld expansion gives
\begin{align}
\psi_a(\omega,T)
 = {} & 
\frac{e^3}
{8\hbar^2(2\pi)^3}
\sum_{\chi }
\chi
\Bigg[
\mathcal P_{a,h}(E,\omega) +
\frac{\pi^2}{6}
(k_BT)^2
\nonumber\\
& \times
\frac{\partial^2
\mathcal P_{a,h}(E,\omega)}
{\partial E^2}
\Bigg]_{E=\mu_\chi} \!\!
+
\mathcal O\!\left[(k_BT)^4\right].
\label{eq:psi_magnetic_sommerfeld}
\end{align}

To quantify the impact of polarized magnetic impurities on the nonlinear Hall response, in Fig. \ref{fig:RectM} we show the numerical evaluation of the longitudinal and transverse rectification efficiencies, $\psi_\parallel(\omega,T)$ and $\psi_\perp(\omega,T)$, governed by Eqs. \eqref{eq:psi_magnetic_zeroT} and \eqref{eq:psi_magnetic_sommerfeld}. Since the contact interaction in
Eq.~\eqref{Magnetic_disorder} requires an integrated coupling with units
of energy times volume, we define
$(JS)_{\rm cont}=\Delta_{\rm ex}V_{\rm cell}$. To parameterize the continuum exchange interaction we use the exchange
energy scale $\Delta_{\rm ex}=165$ meV employed to describe the spin-splitting induced gap observed in EuCd$_2$As$_2$ \cite{Jo_NatC2021}. Using the lattice parameters of EuCd$_2$As$_2$, the hexagonal primitive-cell
volume is $V_{\rm cell}=(\sqrt{3}/2)a^2c\simeq0.125$ nm$^3$
\cite{Wang_PRB2019}, giving
$(JS)_{\rm cont}\simeq20.6$ meV\,nm$^3$. This procedure is a
phenomenological matching of energy scales and should not be interpreted as a microscopic determination of $J$ for the present model. Rather, we intend to illustrate the transport physics of magnetic scattering through measured values in the representative EuCd$_2$As$_2$. All other parameters remain identical to those used previously. As established analytically in Eq.~\eqref{eq:xi_magnetic_decomposition}, the presence of a preferred magnetization axis $\hat{m}$ breaks the rotational symmetry of the scattering process. However, as illustrated in Figs. \ref{fig:Rect_Mag_diff} and \ref{fig:RectT_Mag_diff}, the anisotropic difference $(\psi_\parallel - \psi_\perp)$ is 3 to 4 orders of magnitude smaller than the individual components. While this difference is strictly finite, its negligible magnitude renders $\psi_\perp$ virtually indistinguishable from $\psi_\parallel$. Magnetic polarization is required in the present calculation to generate the helicity-dependent anisotropic correction and the finite second-harmonic tensor structure. It is not, however, sufficient to produce a nonzero net anomalous response when the two Weyl nodes are
energetically equivalent.

The previous results enable us to evaluate the nonlinear currents. The tensor structure in Eq.~(\ref{eq:xi_magnetic_decomposition}) leads to a rectified anomalous current
\begin{align}
\mathbf J_{\rm anom}^{(0)}
=
2 \operatorname{Re}
\Big[
&
\psi_\perp
\boldsymbol{\mathcal E}
\times
\boldsymbol{\mathcal E}^{*}
+
(
\psi_\parallel-\psi_\perp )
(
\hat{\mathbf m}\cdot\boldsymbol{\mathcal E}
) (
\hat{\mathbf m}
\times
\boldsymbol{\mathcal E}^{*} )
\Big] .
\label{eq:Jdc_magnetic}
\end{align}
The first term has the same structure as in the scalar case, whereas the second term is induced entirely by the directional anisotropy of the magnetic scattering. It depends explicitly on the relative orientation between the optical field and the impurity polarization.

For the second-harmonic response, the isotropic term proportional to $\epsilon_{ijk}$ vanishes because $\boldsymbol{\mathcal E}\times\boldsymbol{\mathcal E}=0$. The anisotropic part, however, remains finite:
\begin{align}
\mathbf J_{\rm anom}^{(2\omega)}
=
2 \operatorname{Re}
\Big[
( \psi_\parallel-\psi_\perp )
( \hat{\mathbf m}\cdot\boldsymbol{\mathcal E} ) (
\hat{\mathbf m}
\times
\boldsymbol{\mathcal E} )
\,
e^{2i\omega t}
\Big].
\label{eq:J2omega_magnetic}
\end{align}
Thus, polarized magnetic disorder generates a finite anomalous second-harmonic current even though the underlying Weyl dispersion remains isotropic. This response vanishes when the electric field is either parallel or perpendicular to $\hat{\mathbf m}$ and is maximal for an intermediate relative orientation.

It is important to note that if the two Weyl nodes are energetically equivalent and otherwise identical, $\mu_+=\mu_-$, their anomalous contributions still cancel in the sum over chirality. Polarized magnetic disorder changes the tensorial structure of the response, but a finite net anomalous conductivity in the symmetric two-node model still requires inequivalent node contributions.

\section{Conclusions and outlook} \label{Conclusion_Section}

In this work we developed a microscopic semiclassical theory describing the combined influence of disorder and breaking of inversion symmetry on the nonlinear Hall response of Weyl semimetals. Using the semiclassical formalism, with the transport relaxation time being estrictly treated as a momentum-dependent quantity, we derived general expresions for both the rectified and the second-harmonic conductivity tensors. This formulation provides a unified framework in which the nonlinear response can be evaluated for different microscopic scattering mechanisms without modifying the transport formalism.

A central result of our analysis is that the geometric contribution, associated with the Berry curvature, can be clearly separated from the microscopic information encoded in the impurity scattering. Within this approach,  all disorder effects enter exclusively through the transport relaxation time, whereas the Berry curvature remains entirely determined by the Bloch states. This separation allows the role of different impurity mechanisms to be identified transparently and provides a flexible starting point for incorporating more realistic scattering models in future studies.

For scalar disorder, including short-range, Gaussian and Coulomb impurities, we found that the relaxation time remains isotropic and chirality independent. In this case, the nonlinear Hall response is entirely controlled by the anomalous Berry-curvature contribution, while the quasiparticle contribution vanishes by rotational symmetry. A finite rectified current emerges only when the two Weyl nodes become energetically inequivalent, preventing the exact cancellation of their opposite Berry-curvature contributions. Different impurity models nevertheless produce quantitatively distinct dependences on the chemical potential and temperature through their characteristic energy-dependent relaxation times.

The magnetic case exhibits a considerably richer structure. We demonstrated that neither the first Born approximation nor the pure second-order contributino of the $T$-matrix expansion generates different transport lifetimes for opposite helicities. Instead, chirality-dependent momentum relaxation emerges only through the interference between first- and second-order scattering amplitudes, producing an anisotropic relaxation time for polarized magnetic impurities. Although this anisotropy modifies the tensorial structure of the nonlinear Hall conductivity abd enables second-harmonic generation, our numerical calculations show that the anisotropic corrections remains several orders of magnitude smaller than the dominant isotropic contribution for representative material parameters.

The formalism developed here opens several directions for future work. An immediate extension consists in considering anisotropic or tilted Weyl cones, for which the quaisparticle contribution is expected to become finite and may compete with the Berry-curvature response. Another natural developments is the incorporation of extrinsic mechanisms such as side-jump and skew-scattering contributions within a quantum kinetic description beyond the relaxation-time approximation. It would also be interesting to include external magnetic fields, where the interplay between Berry curvature, orbital magnetic moments, and disorder is expected to generate additional nonlinear magnetotransport phenomena. More broadly, since our formulation relies only on the semiclassical band geometry and the microscopic transport lifetime, it can be readily generalized to other topological semimetals, including Dirac, multifold, and nodal-line systems, where the role of disorder in nonlinear transport remains largely unexplored.

\acknowledgements{J.A.C. gratefully acknowledges the support of SECIHTI through the program \textit{Becas Nacionales para estudios de Posgrado}, under grant number 4018746. D.A.B. was supported by the DGAPA-UNAM Posdoctoral Program. A.M.-R. acknowledges financial support by UNAM-PAPIIT project No. IG100224, UNAM-PAPIME project No. PE109226, by SECIHTI project No. CBF-2025-I-1862 and by the Marcos Moshinsky Foundation}

\appendix

\section{Magnetic-impurity scattering amplitude in the helicity basis} \label{Appendix_Scatt_Amplitude_1er_order}

Here we compute explicitly the spinor matrix element
\begin{align}
    \mathbf{V} \equiv \bra{\mathbf{k} ', h '} \boldsymbol{\sigma} \ket{\mathbf{k}, h } . 
\end{align}
To this end, it is useful to recall the following vectorial identities of Pauli matrices:
\begin{align}
    \boldsymbol{\sigma} (\boldsymbol{\sigma} \cdot \hat{\mathbf{k}})  = \hat{\mathbf{k}} + i \boldsymbol{\sigma} \times \hat{\mathbf{k}} , \quad (\boldsymbol{\sigma} \cdot \hat{\mathbf{k}}')  \boldsymbol{\sigma} = \hat{\mathbf{k}}' - i \boldsymbol{\sigma} \times \hat{\mathbf{k}}' .
\end{align}
In the helicity basis the spinors satisfy
\begin{align}
    (\boldsymbol{\sigma} \cdot \hat{\mathbf{k}}) \ket{\mathbf{k}, h } = h \ket{\mathbf{k}, h }, \quad (\boldsymbol{\sigma} \cdot \hat{\mathbf{k}}') \ket{\mathbf{k}', h' } = h' \ket{\mathbf{k}', h' } .
\end{align}
which allows us to rewrite the matrix element as
\begin{align}
     \bra{\mathbf{k}', h' } \boldsymbol{\sigma} \ket{\mathbf{k}, h }  &= h \bra{\mathbf{k}', h' }  (  \hat{\mathbf{k}} + i \boldsymbol{\sigma} \times \hat{\mathbf{k}} ) \ket{\mathbf{k}, h }  . 
\end{align}
From this expression one obtains the relation
\begin{align}
    \mathbf{V} = h \left[ \mathcal{A} _{hh'} (\theta _{\mathbf{k}\mathbf{k}'}) \, \hat{\mathbf{k}}  + i \mathbf{V} \times \hat{\mathbf{k}} \right] , \label{Aux1}
\end{align}
where the overlap between initial and final spinors is given by
\begin{align}
    \mathcal{A} _{hh'} (\theta _{\mathbf{k}\mathbf{k}'}) = \langle \mathbf{k}', h' \vert \mathbf{k}, h \rangle = \sqrt{\frac{1 + hh' \cos \theta _{\mathbf{k}\mathbf{k}'} }{2}} .  \label{Overlap}
\end{align}
A completely analogous calculation starting from the primed state leads to
\begin{align}
    \mathbf{V} = h' \left[ \mathcal{A} _{hh'} (\theta _{\mathbf{k}\mathbf{k}'}) \, \hat{\mathbf{k}}' - i \mathbf{V} \times \hat{\mathbf{k}}' \right] . \label{Aux2}
\end{align}
These two equivalent forms yield the scalar projections
\begin{align}
    \hat{\mathbf{k}} \cdot \mathbf{V} = h \, \mathcal{A} _{hh'} (\theta _{\mathbf{k}\mathbf{k}'}) , \quad  \hat{\mathbf{k}}' \cdot \mathbf{V} = h' \, \mathcal{A} _{hh'} (\theta _{\mathbf{k}\mathbf{k}'}) .  
\end{align}
Taking the cross product of Eq. (\ref{Aux2}) with $\hat{\mathbf{k}}'$ and using the scalar projections, one finds the identity
\begin{align}
     \hat{\mathbf{k}}' \times \mathbf{V} = i \mathcal{A} _{hh'} (\theta _{\mathbf{k}\mathbf{k}'}) \, \hat{\mathbf{k}}' - i h' \mathbf{V}  .
\end{align}
Substituting Eq. (\ref{Aux1}) into the previous result, the matrix element can finally be written in compact closed form as
\begin{align}
      \mathbf{V} &= \frac{\mathcal{A} _{hh'} (\theta _{\mathbf{k}\mathbf{k}'})}{ 1 + h h' \cos \theta _{\mathbf{k}\mathbf{k}'} } \left[ h   \hat{\mathbf{k}}    + h' \hat{\mathbf{k}}' +   i  h h' \hat{\mathbf{k}}' \times \hat{\mathbf{k}} \right] .  \label{V_final}
\end{align}
This closed expression is the result quoted in the main text in Eq. (\ref{Scatt_Magnetic_1order}), and makes explicit the dependence of the scattering vertex on the helicities of the incoming and outgoing states.

\section{Disorder-averaged squared matrix element} \label{Disorder_averaged_matrix_element}

In this Appendix we compute the squared matrix element of Eq.~\eqref{Scatt_Magnetic_1order} and perform the spatial average over randomly distributed impurities. This procedure yields the disorder-averaged transition probability that enters the Fermi golden rule and determines the relaxation rates discussed in the main text.

From Eq.~\eqref{Scatt_Magnetic_1order} we define, for brevity, the real vectors
\begin{align}
    \mathbf{I} = h   \hat{\mathbf{k}}    + h' \hat{\mathbf{k}}' , \qquad \mathbf{J} =  h h' \hat{\mathbf{k}}' \times \hat{\mathbf{k}} . 
\end{align}
so that the scattering amplitude (\ref{Scatt_Magnetic_1order}) can be expressed in the compact form
\begin{align}
 \mathcal{M}^{(1)} _{\mathbf{k}'h', \mathbf{k}h} = \frac{J}{2   \mathcal{A} _{hh'} (\theta _{\mathbf{k}\mathbf{k}'})}   \tilde{\mathbf{S}} (\mathbf{q}) \cdot \left( \mathbf{I} +   i  \mathbf{J} \right) .   
\end{align}
The squared modulus is then simply the modulus of the complex scalar $\tilde{\mathbf{S}} (\mathbf{q}) \cdot \left( \mathbf{I} +   i  \mathbf{J} \right)$, which gives
\begin{align}
  \vert \mathcal{M}^{(1)} _{\mathbf{k}'h', \mathbf{k}h} \vert ^{2} &= \frac{J ^{2}}{4   \mathcal{A} _{hh'} ^{2} (\theta _{\mathbf{k}\mathbf{k}'})} \Big\{\ \!\! \vert \tilde{\mathbf{S}} \cdot  \mathbf{I} \vert ^{2} + \vert \tilde{\mathbf{S}}  \cdot  \mathbf{J} \vert ^{2} \notag \\ & \hspace{2cm} + 2 \, \mbox{Im} \, [  (\tilde{\mathbf{S}} \cdot \mathbf{I} ) (\tilde{\mathbf{S}} \cdot \mathbf{J} ) ^{\ast}   ]  \Big\}\ ,
\end{align}
where we suppressed the $\mathbf{q}$ dependence of $\tilde{\mathbf{S}}$ for brevity. This relation can also be written in component form, which is convenient when performing the disorder average. Introducing Cartesian indices $i,j=x,y,z$ one finds
\begin{align}
  \vert \mathcal{M}^{(1)} _{\mathbf{k}'h', \mathbf{k}h} \vert ^{2} &= \frac{J ^{2}}{4   \mathcal{A} _{hh'} ^{2} (\theta _{\mathbf{k}\mathbf{k}'})} \Big\{\ \!\!  \tilde{S} _{i} \tilde{S} ^{\ast} _{j} \, (I _{i} I _{j} + J _{i} J _{j})   \notag \\ & \hspace{2.5cm} + 2 I _{i} J _{j}  \, \mbox{Im} \, ( \tilde{S} _{i}  \tilde{S} ^{\ast} _{j}  ) \Big\}\ ,
\end{align}
with implicit summation over repeated indices. For the disorder average we assume uncorrelated, isotropically oriented local moments and random, independent impurity positions. In that case the spin-density correlator in momentum space is diagonal and real,
\begin{align}
    \langle \tilde{S} _{i} (\mathbf{q}) \tilde{S} ^{\ast} _{j} (\mathbf{q}) \rangle = \frac{1}{3} S(S+1) \, n _{i} \, \delta _{ij} , 
\end{align}
where $n _{i}$ is the impurity density (the usual volume factors are absorbed by the continuum normalization of states). Because the correlator is real and proportional to $\delta _{ij}$, the antisymmetric term proportional to $\mbox{Im} \, ( \tilde{S} _{i}  \tilde{S} ^{\ast} _{j}  )$ vanishes upon averaging, and only the symmetric contraction survives:
\begin{align}
  \langle \vert \mathcal{M}^{(1)} _{\mathbf{k}'h', \mathbf{k}h} \vert ^{2} \rangle &= \frac{n _{i} J ^{2} S(S+1)}{12   \mathcal{A} _{hh'} ^{2} (\theta _{\mathbf{k}\mathbf{k}'})} \; (\mathbf{I} \cdot \mathbf{I} + \mathbf{J} \cdot \mathbf{J} ) .   
\end{align}
Finally, using
\begin{align}
    \mathbf{I} \cdot \mathbf{I} = 2 (1+hh' \cos \theta _{\mathbf{k}\mathbf{k}'} ) , \quad \mathbf{J} \cdot \mathbf{J} = \sin ^{2} \theta _{\mathbf{k}\mathbf{k}'}  ,
\end{align}
we establish the result of Eq. (\ref{Average_Mag_Impurity}) of the main text. This is the disorder-averaged squared matrix element that enters the Fermi golden rule.

\section{Scattering from polarized magnetic impurities}
\label{app:polarized_magnetic_impurities}

In this Appendix we derive the scattering probability produced by polarized magnetic impurities given by Eq. (\ref{M1_unpolarized}) and show explicitly that the resulting transport relaxation time is anisotropic, but independent of the sign of the incoming helicity. We consider a local magnetic impurity potential of the form
\begin{align}
    V(\mathbf{r}) = J S \, \hat{\mathbf{m}} \cdot \boldsymbol{\sigma} \, \delta ( \mathbf{r} ) , \label{app:polarized_potential}
\end{align}
where $J$ is the exchange coupling, $S$ is the magnitude of the localized magnetic moment, and $\hat{\mathbf{m}}$ is the corresponding polarization axis. Within the first Born approximation, in the helicity bases, the scattering amplitude from the
state $|\mathbf k,h\rangle$ to $|\mathbf k',h'\rangle$ is
\begin{align}
    \mathcal M^{(1)}_{\mathbf k'h',\mathbf kh}
= JS \, \hat{\mathbf m}\cdot \mathbf{V}  ,
\end{align}
where $\mathbf{V} = \bra{\mathbf{k} ', h '} \boldsymbol{\sigma} \ket{\mathbf{k}, h } $ is the matrix element evaluated in the Appendix \ref{Appendix_Scatt_Amplitude_1er_order}, with the final result given by Eq. (\ref{V_final}).

Since all vectors appearing in this expression ($\mathbf{k}$, $\mathbf{k}'$, $\mathbf{m}$) are real, the squared modulus can be separated into symmetric and antisymmetric pieces. The cross term between the real vector $h\hat{\mathbf k}+h'\hat{\mathbf k}'$ and the imaginary vector $ihh'(\hat{\mathbf k}'\times\hat{\mathbf k})$ vanishes. Thus,
\begin{align}
\left| \mathcal M^{(1)}_{\mathbf k'h',\mathbf kh} \right| ^{2} =  \frac{J ^{2} S ^{2}}{2 \left( 1 + h h' \cos \theta \right) }  \left[ ( h \, a + h' \, b )  ^{2}  +  c ^{2}  \right] , \label{app:squared_amplitude_helicity}
\end{align}
where we introduced
\begin{equation}
a = \hat{\mathbf{m}} \cdot \hat{\mathbf{k}} , \qquad b = \hat{\mathbf{m}} \cdot \hat{\mathbf{k}} ' , \qquad c = \hat{\mathbf{m}} \cdot ( \hat{\mathbf{k}} ' \times \hat{\mathbf{k}} ) . \label{app:abc_definition}
\end{equation}
The physical transition rate entering the Boltzmann equation requires a sum over the outgoing helicity. We find
\begin{align}
\sum _{h' = \pm 1} \left| \mathcal M^{(1)}_{\mathbf k'h',\mathbf kh} \right| ^{2} = {} & \frac{J ^{2} S ^{2}}{2} \left[ \frac{ (a+hb) ^{2} + c ^{2}}{ 1 + h \cos \theta } + \frac{ (a-hb) ^{2} + c ^{2} }{ 1 - h \cos \theta } \right] , \label{app:helicity_sum_intermediate}
\end{align}
and combining the two fractions we obtain
\begin{align}
\sum _{h' = \pm 1} \left| \mathcal M^{(1)}_{\mathbf k'h',\mathbf kh} \right| ^{2} = {} & \frac{J ^{2} S ^{2}}{1 - \cos ^{2} \theta } \left[ a ^{2} + b ^{2} + c ^{2} - 2 a b \cos \theta \right] . \label{app:helicity_sum_final}
\end{align}
Restoring the vector notation according to the definitions (\ref{app:abc_definition}) we establish Eq. (\ref{M1_unpolarized}).

Equation~(\ref{app:helicity_sum_final}) is independent of the incoming helicity $h$. Therefore, although the scattering probability is anisotropic because it depends on the orientation of both $\hat{\mathbf{k}}$ and $\hat{\mathbf{k}}'$ relative to $\hat{\mathbf{m}}$, it is identical for the two incoming helicities.

For a random distribution of polarized impurities with density $n_i$, the corresponding transport relaxation rate (\ref{relaxation_time}) becomes
\begin{align}
\frac{1}{\tau _{ h \mathbf{k}} }
={}&
\frac{2\pi n_i J^2S^2}{\hbar}
\nu(E)
\int
\frac{d\Omega_{\mathbf k'}}{4\pi}
\,
\frac{a ^{2} + b ^{2} + c ^{2} - 2 a b \cos \theta }
{1 + \cos \theta} , \label{app:polarized_transport_integral}
\end{align}
where we have used the on-shell condition and introduced the density of states $\nu(E)$.

The integrand in Eq.~(\ref{app:polarized_transport_integral}) depends on the incoming direction through the scalar $\hat{\mathbf m}\cdot\hat{\mathbf k}$, demonstrating that the transport time is anisotropic. Nevertheless, because Eq.~(\ref{app:polarized_transport_integral}) contains no dependence on the sign of $h$, one finds
\begin{align}
    \frac{1}{\tau _{ + \mathbf{k}} } = \frac{1}{\tau _{ - \mathbf{k}} } \equiv \frac{1}{\tau _{ \mathbf{k}} } . 
\end{align}
Thus, polarized magnetic impurities define a preferred axis and generate direction-dependent momentum relaxation, but they do not lift the degeneracy between the scalar transport lifetimes of opposite helicities. The anisotropy must therefore be retained in the solution of the Boltzmann equation, while the relaxation time remains helicity independent.

\section{Second-order amplitude for polarized magnetic impurities} \label{Appendix_M2_polarized}

In this Appendix we derive the expression (\ref{M2}) and (\ref{M2_summed}). We start from the definition of the second-order $T$-matrix element,
\begin{align}
\mathcal{M}^{(2)}_{\mathbf k'h',\mathbf k h}
= \bra{\mathbf k',h'}\,V(\mathbf r)\,G(E)\,V(\mathbf r)\,\ket{\mathbf k,h},
\end{align}
with the exchange potential
\begin{align}
V(\mathbf r) = J\,\mathbf S\cdot\boldsymbol\sigma,
\end{align}
and the local Green's function $G _{\chi} (E) = G (E) \, \sigma _{0}$ of the clean Weyl Hamiltonian, which is proportional to the identity in spinor space, cf. Eq.~(\ref{Green}). For a polarized impurity with fixed moment
$\mathbf S = S\,\hat{\mathbf m}$, one has
\begin{align}
V(\mathbf r) = J S\,(\hat{\mathbf m}\cdot\boldsymbol\sigma).
\end{align}
Inserting this into the expression for $\mathcal M^{(2)}$ and using that
$G(E)\propto\sigma_{0}$ commutes with Pauli matrices, we obtain
\begin{align}
\mathcal{M}^{(2)}_{\mathbf k'h',\mathbf k h}
&= J^{2} S^{2}\,G(E)\,
\bra{\mathbf k',h'}\,
(\hat{\mathbf m}\cdot\boldsymbol\sigma)\,
(\hat{\mathbf m}\cdot\boldsymbol\sigma)\,
\ket{\mathbf k,h}.
\end{align}
The spin structure simplifies using the identity $(\hat{\mathbf m}\cdot\boldsymbol\sigma)^{2} = \sigma_{0}$, where we used $\hat{\mathbf m}$ as a unit vector. Therefore,
\begin{align}
\mathcal{M}^{(2)}_{\mathbf k'h',\mathbf k h}
&=  J^{2} S^{2}\,G(E)\,
\langle \mathbf k',h'|\mathbf k,h\rangle.
\end{align}
By definition, the helicity spinor overlap $\mathcal A_{hh'}(\theta_{\mathbf k\mathbf k'})
\equiv \langle \mathbf k',h'|\mathbf k,h\rangle$ is given by Eq. (\ref{Overlap}), so that the second-order amplitude can be written as
\begin{align}
\mathcal{M}^{(2)}_{\mathbf k'h',\mathbf k h}
= J^{2} S^{2}\,G(E)\,\mathcal A_{hh'}(\theta_{\mathbf k\mathbf k'}) ,
\end{align}
which corresponds to Eq. (\ref{M2}). Taking the modulus squared and summing over the outgoing helicity $h'$ gives
\begin{align}
\sum_{h'} \big|\mathcal{M}^{(2)}_{\mathbf k'h',\mathbf k h}\big|^{2}
&= J^{4} S^{4}\,|G(E)|^{2}
\sum_{h'} \big|\mathcal A_{hh'}(\theta_{\mathbf k\mathbf k'})\big|^{2}.
\end{align}
Using the explicit form of the overlap,
\begin{align}
\big|\mathcal A_{hh'}(\theta)\big|^{2}
= \frac{1+h h'\cos\theta}{2},
\end{align}
we obtain
\begin{align}
\sum_{h'=\pm 1} \big|\mathcal A_{hh'}(\theta)\big|^{2}
= \frac{1+h\cos\theta}{2} + \frac{1-h\cos\theta}{2}
= 1.
\end{align}
Therefore,
\begin{align}
\sum_{h'} \big|\mathcal{M}^{(2)}_{\mathbf k'h',\mathbf k h}\big|^{2}
= J^{4} S^{4}\,|G(E)|^{2}, \label{app:helicity_sum_final_2}
\end{align}
which proves the result presented in Eq. (\ref{M2_summed}).

\section{Interference term for polarized magnetic impurities} \label{app:polarized}

Here we evaluate the interference contribution between the first- and second-order Born amplitudes for polarized magnetic impurities appearing in Eq. (\ref{second_order_Born_app}). The calculation is performed directly at the operator level, avoiding any explicit manipulation of the gauge dependent phases of the helicity spinors.

Our starting points are the first- and second-order amplitudes given by Eqs. (\ref{M1_amplitude}) and (\ref{Scatt_Magnetic_2order}), respectively. Using these expressions, the interference contribution to the scattering probability becomes
\begin{align}
\Big\langle \mathcal M^{(1)}_{\mathbf k'h',\mathbf kh}
\mathcal M^{(2)*}_{\mathbf k'h',\mathbf kh} \Big\rangle
= {} &
J^3S^3G^*(E)
\left\langle
\mathbf k',h'
\left|
\hat{\mathbf m}\cdot\boldsymbol{\sigma}
\right|
\mathbf k,h
\right\rangle
\nonumber\\
&\hspace{1.5cm}\times
\left\langle
\mathbf k,h
\middle|
\mathbf k',h'
\right\rangle .
\label{app:interference_before_completeness}
\end{align}
The two matrix elements appearing in each term of the sum are ordinary complex numbers and can therefore be reordered. Thus, using the completeness relation for the two helicity states at fixed outgoing momentum $\mathbf k'$, $\sum_{h'=\pm1}
|\mathbf k',h'\rangle
\langle\mathbf k',h'|
=
\sigma_0$, we get
\begin{align}
\sum_{h'}
\Big\langle \mathcal M^{(1)}_{\mathbf k'h',\mathbf kh}
\mathcal M^{(2)*}_{\mathbf k'h',\mathbf kh} \Big\rangle
={}&
J^3S^3G^*(E)
\left\langle
\mathbf k,h
\left|
\hat{\mathbf m}\cdot\boldsymbol{\sigma}
\right|
\mathbf k,h
\right\rangle .
\label{app:interference_expectation}
\end{align}
Since the incoming state is an eigenstate of the helicity operator, $(
\boldsymbol{\sigma}\cdot\hat{\mathbf k})
|\mathbf k,h\rangle
=
h|\mathbf k,h\rangle$, and its pseudospin expectation value is $\left\langle
\mathbf k,h
\left|
\boldsymbol{\sigma}
\right|
\mathbf k,h
\right\rangle
=
h\hat{\mathbf k}$, we find $\left\langle
\mathbf k,h
\left|
\hat{\mathbf m}\cdot\boldsymbol{\sigma}
\right|
\mathbf k,h
\right\rangle
=
h\,
\hat{\mathbf m}\cdot\hat{\mathbf k}$. Therefore, Eq. (\ref{app:interference_expectation}) becomes
\begin{align}
\sum_{h'}
\Big\langle \mathcal M^{(1)}_{\mathbf k'h',\mathbf kh}
\mathcal M^{(2)*}_{\mathbf k'h',\mathbf kh} \Big\rangle
={}&
h J^3S^3G^*(E) \, (\hat{\mathbf m}\cdot\hat{\mathbf k})  .
\label{app:interference_complex_result}
\end{align}
Taking the real part we obtain Eq. (\ref{eq:interference_result}). This result has several relevant properties. First, it is manifestly independent of the gauge chosen for the helicity spinors. Second, it contains no dependence on the outgoing direction $\hat{\mathbf k}'$. Third, it selects the real principal-value part $C_\Lambda(E)$ of the local Green function rather than its imaginary on-shell component. Therefore, no skew-scattering term survives after the complete sum over the outgoing helicity.

The complete transport relaxation rate up to second order in the $T$-matrix is obtained from
\begin{align}
    \frac{1}{\tau _{ h \mathbf{k}} } = \frac{2\pi n_i}{\hbar} \nu (E) \int \frac{d \Omega _{\mathbf{k}'}}{4 \pi} \sum _{h'} \left| \mathcal{M} _{\mathbf{k}'h',\mathbf{k}h} \right| ^{2} ( 1 - \hat{\mathbf{k}} \cdot \hat{\mathbf{k}}' ) , \label{app:complete_transport_rate}
\end{align}
where, up to the order considered, 
\begin{align}
    \vert \mathcal{M} _{\mathbf{k}'h', \mathbf{k}h} \vert ^{2} &= \vert \mathcal{M} ^{(1)}  _{\mathbf{k}'h', \mathbf{k}h} \vert ^{2} + \vert \mathcal{M} ^{(2)}  _{\mathbf{k}'h', \mathbf{k}h} \vert ^{2} \notag \\[5pt] & \hspace{1.5cm} + 2 \, \mbox{Re} \left[ \mathcal{M} ^{(1)}  _{\mathbf{k}'h', \mathbf{k}h} \, \mathcal{M} ^{(2) \ast} _{\mathbf{k}'h', \mathbf{k}h} \right] . \label{second_order_Born_app_2}
\end{align}
The three contributions entering Eq.~(\ref{app:complete_transport_rate}), namely the first-order term, the second-order term and the interference contribution [in ther order they appear in Eq. (\ref{second_order_Born_app_2})], were evaluated separately. The final results are given by Eqs. (\ref{app:helicity_sum_final}), (\ref{app:helicity_sum_final_2}) and (\ref{app:interference_complex_result}), respectively. The angular integrals required by Eq.~(\ref{app:complete_transport_rate}) are simple. The results are
\begin{align}
\int \frac{d \Omega_{\mathbf{k}'}}{4 \pi} \sum _{h'} \left| \mathcal M ^{(1)} _{\mathbf{k} 'h',\mathbf{k}h} \right| ^{2} ( 1 - \hat{\mathbf{k}} \cdot \hat{\mathbf{k}}' ) &= J ^{2} S ^{2} , \\ \int \frac{d \Omega_{\mathbf{k}'}}{4 \pi} \sum _{h'} \left| \mathcal M ^{(2)} _{\mathbf{k} 'h',\mathbf{k}h} \right| ^{2} ( 1 - \hat{\mathbf{k}} \cdot \hat{\mathbf{k}}' ) &= J ^{4} S ^{4} |G(E)| ^{2} ,
\end{align}
and
\begin{align}
\int \frac{d \Omega_{\mathbf{k}'}}{4 \pi} \sum _{h'} 2 \, \mbox{Re} \left[ \mathcal{M} ^{(1)}  _{\mathbf{k}'h', \mathbf{k}h} \, \mathcal{M} ^{(2) \ast} _{\mathbf{k}'h', \mathbf{k}h} \right] ( 1 - \hat{\mathbf{k}} \cdot \hat{\mathbf{k}}' ) \notag \\  = 2hJ ^{3} S ^{3} C _{\Lambda} (E) \, ( \hat{\mathbf{m}} \cdot \hat{\mathbf{k}} ) . 
\end{align}
Collecting these terms, the complete transport relaxation rate becomes
\begin{align}
    \frac{1}{\tau _{ h \mathbf{k}} } &= \frac{2\pi n_i J ^{2} S ^{2} }{\hbar} \, \nu (E) \, \Big[ 1 + J ^{2} S ^{2} |G(E)| ^{2} \notag \\ & \hspace{3.3cm}  +  2hJ S C _{\Lambda} (E) \, ( \hat{\mathbf{m}} \cdot \hat{\mathbf{k}} ) \Big] . 
\end{align}
Thus, the first- and second-order squared amplitudes determine the
helicity-even part of the relaxation rate, while their interference
produces the anisotropic helicity-odd contribution proportional to
$h\,\hat{\mathbf m}\cdot\hat{\mathbf k}$.

\bibliography{PHE.bib}

@article{PhysRevB.103.035102,
  title = {Quantized electrochemical transport in Weyl semimetals},
  author = {Flores-Calder\'on, R. and Mart\'{\i}n-Ruiz, A.},
  journal = {Phys. Rev. B},
  volume = {103},
  issue = {3},
  pages = {035102},
  numpages = {12},
  year = {2021},
  month = {Jan},
  publisher = {American Physical Society},
  doi = {10.1103/PhysRevB.103.035102},
  url = {https://link.aps.org/doi/10.1103/PhysRevB.103.035102}
}

@article{PhysRevB.108.155132,
  title = {Planar Hall effect in Weyl semimetals induced by pseudoelectromagnetic fields},
  author = {Medel Onofre, L. and Mart\'{\i}n-Ruiz, A.},
  journal = {Phys. Rev. B},
  volume = {108},
  issue = {15},
  pages = {155132},
  numpages = {16},
  year = {2023},
  month = {Oct},
  publisher = {American Physical Society},
  doi = {10.1103/PhysRevB.108.155132},
  url = {https://link.aps.org/doi/10.1103/PhysRevB.108.155132}
}

@article{sci_reports_10.1038,
	author = {Medel, Leonardo and Ghosh, Rahul and Mart{\'\i}n-Ruiz, Alberto and Mandal, Ipsita},
	date = {2024/09/13},
	doi = {10.1038/s41598-024-68615-0},
	id = {Medel2024},
	isbn = {2045-2322},
	journal = {Scientific Reports},
	number = {1},
	pages = {21390},
	title = {Electric, thermal, and thermoelectric magnetoconductivity for Weyl/multi-Weyl semimetals in planar Hall set-ups induced by the combined effects of topology and strain},
	url = {https://doi.org/10.1038/s41598-024-68615-0},
	volume = {14},
	year = {2024}}

@article{RevModPhys.90.015001,
  title = {{W}eyl and {D}irac semimetals in three-dimensional solids},
  author = {Armitage, N. P. and Mele, E. J. and Vishwanath, A.},
  journal = {Rev. Mod. Phys.},
  volume = {90},
  issue = {1},
  pages = {015001},
  numpages = {57},
  year = {2018},
  month = {Jan},
  publisher = {American Physical Society},
  doi = {10.1103/RevModPhys.90.015001},
  url = {https://link.aps.org/doi/10.1103/RevModPhys.90.015001}
}

@article{RevModPhys.82.1959,
  title = {Berry phase effects on electronic properties},
  author = {Xiao, D. and Chang, M.-C. and Niu, Q.},
  journal = {Rev. Mod. Phys.},
  volume = {82},
  issue = {3},
  pages = {1959--2007},
  numpages = {0},
  year = {2010},
  month = {Jul},
  publisher = {American Physical Society},
  doi = {10.1103/RevModPhys.82.1959},
  url = {https://link.aps.org/doi/10.1103/RevModPhys.82.1959}
}

@article{PhysRevB.59.14915,
  title = {Wave-packet dynamics in slowly perturbed crystals: Gradient corrections and {B}erry-phase effects},
  author = {Sundaram, G. and Niu, Q.},
  journal = {Phys. Rev. B},
  volume = {59},
  issue = {23},
  pages = {14915--14925},
  numpages = {0},
  year = {1999},
  month = {Jun},
  publisher = {American Physical Society},
  doi = {10.1103/PhysRevB.59.14915},
  url = {https://link.aps.org/doi/10.1103/PhysRevB.59.14915}
}

@article{BonillaNano2021,
  author  = {Bonilla, Daniel and Mu{\~n}oz, Enrique and Soto-Garrido, Rodrigo},
  title   = {{Thermo-Magneto-Electric Transport through a Torsion Dislocation in a Type I Weyl Semimetal}},
  journal = {Nanomaterials},
  volume  = {11},
  number  = {11},
  pages   = {2972},
  year    = {2021},
  doi     = {10.3390/nano11112972}
}

@article{BonillaNano2022,
  author  = {Bonilla, Daniel and Mu{\~n}oz, Enrique},
  title   = {{Electronic Transport in Weyl Semimetals with a Uniform Concentration of Torsional Dislocations}},
  journal = {Nanomaterials},
  volume  = {12},
  number  = {20},
  pages   = {3711},
  year    = {2022},
  doi     = {10.3390/nano12203711}
}

@article{BonillaNA2024,
  author  = {Bonilla, Daniel A. and Mu{\~n}oz, Enrique},
  title   = {{Thermoelectric transport in Weyl semimetals under a uniform concentration of torsional dislocations}},
  journal = {Nanoscale Adv.},
  volume  = {6},
  pages   = {2701--2712},
  year    = {2024},
  doi     = {10.1039/d4na00056k}
}

@article{ParraPRB2026,
  title = {{Electron-phonon interactions and instabilities in Weyl semimetals under magnetic fields and torsional strain}},
  author = {Parra, Fabi\'an Jofr\'e and Bonilla, Daniel A. and Mu\~noz, Enrique},
  journal = {Phys. Rev. B},
  volume = {113},
  issue = {7},
  pages = {075109},
  numpages = {34},
  year = {2026},
  month = {Feb},
  publisher = {American Physical Society},
  doi = {10.1103/9qz1-2wwd},
  url = {https://link.aps.org/doi/10.1103/9qz1-2wwd}
}

@article{LIU2024101343,
title = {Recent progress in topological semimetal and its realization in Heusler compounds},
journal = {Materials Today Physics},
volume = {41},
pages = {101343},
year = {2024},
issn = {2542-5293},
doi = {https://doi.org/10.1016/j.mtphys.2024.101343},
url = {https://www.sciencedirect.com/science/article/pii/S2542529324000191},
author = {Hongshuang Liu and Jiashuo Liang and Taiyu Sun and Liying Wang}
}

@Article{Deng2019,
author={Deng, Tao
and Yang, Hai-Feng
and Zhang, Jing
and Li, Yi-Wei
and Yang, Le-Xian
and Liu, Zhong-Kai
and Chen, Yu-Lin},
title={Progress of {ARPES} study on topological semimetals},
journal={Acta Physica Sinica},
year={2019},
volume={68},
number={22},
pages={227102-1-227102-22},
doi={10.7498/aps.68.20191544},
url={https://doi.org/10.7498/aps.68.20191544}
}

@article{Bernevig2018,
author = {Bernevig ,Andrei and Weng ,Hongming and Fang ,Zhong and Dai ,Xi},
title = {Recent Progress in the Study of Topological Semimetals},
journal = {Journal of the Physical Society of Japan},
volume = {87},
number = {4},
pages = {041001},
year = {2018},
doi = {10.7566/JPSJ.87.041001},
URL = {https://doi.org/10.7566/JPSJ.87.041001}
}

@Article{Zou2019,
author={Zou, Jinyu
and He, Zhuoran
and Xu, Gang},
title={The study of magnetic topological semimetals by first principles calculations},
journal={npj Computational Materials},
year={2019},
month={Oct},
day={04},
volume={5},
number={1},
pages={96},
issn={2057-3960},
doi={10.1038/s41524-019-0237-5},
url={https://doi.org/10.1038/s41524-019-0237-5}
}

@Article{Schoop2020,
author={Schoop, Leslie M.
and Dai, Xi
and Cava, R. J.
and Ilan, Roni},
title={Special topic on topological semimetals---New directions},
journal={APL Materials},
year={2020},
month={Mar},
day={23},
volume={8},
number={3},
pages={030401},
issn={2166-532X},
doi={10.1063/5.0006015},
url={https://doi.org/10.1063/5.0006015}
}

@Article{Bernevig2022,
author={Bernevig, B. Andrei
and Felser, Claudia
and Beidenkopf, Haim},
title={Progress and prospects in magnetic topological materials},
journal={Nature},
year={2022},
month={Mar},
day={01},
volume={603},
number={7899},
pages={41-51},
issn={1476-4687},
doi={10.1038/s41586-021-04105-x},
url={https://doi.org/10.1038/s41586-021-04105-x}
}

@article{YuAFM2025,
author = {Yu, Huihui and Zeng, Haoran and Zhang, Yanzhe and Liu, Yihe and ShangGuan, Wei and Zhang, Xiankun and Zhang, Zheng and Zhang, Yue},
title = {Two-Dimensional Layered Topological Semimetals for Advanced Electronics and Optoelectronics},
journal = {Advanced Functional Materials},
volume = {35},
number = {2},
pages = {2412913},
doi = {https://doi.org/10.1002/adfm.202412913},
url = {https://advanced.onlinelibrary.wiley.com/doi/abs/10.1002/adfm.202412913},
year = {2025}
}

@article{ZhongAM2025,
author = {Zhong, Mengyuan and Vu, Nam Thanh Trung and Zhai, Wenhao and Soh, Jian Rui and Liu, Yuanda and Wu, Jing and Suwardi, Ady and Liu, Huajun and Chang, Guoqing and Loh, Kian Ping and Gao, Weibo and Qiu, Cheng-Wei and Yang, Joel K. W. and Dong, Zhaogang},
title = {{W}eyl Semimetals: From Principles, Materials to Applications},
journal = {Advanced Materials},
volume = {37},
number = {37},
pages = {2506236},
doi = {https://doi.org/10.1002/adma.202506236},
url = {https://advanced.onlinelibrary.wiley.com/doi/abs/10.1002/adma.202506236},
year = {2025}
}

@article{JinPRL2020,
  title = {Two-Dimensional {D}irac Semimetals without Inversion Symmetry},
  author = {Jin, Y. J. and Zheng, B. B. and Xiao, X. L. and Chen, Z. J. and Xu, Y. and Xu, H.},
  journal = {Phys. Rev. Lett.},
  volume = {125},
  issue = {11},
  pages = {116402},
  numpages = {5},
  year = {2020},
  month = {Sep},
  publisher = {American Physical Society},
  doi = {10.1103/PhysRevLett.125.116402},
  url = {https://link.aps.org/doi/10.1103/PhysRevLett.125.116402}
}

@article{YanARCMP2017,
   author = "Yan, Binghai and Felser, Claudia",
   title = "Topological Materials: {W}eyl Semimetals", 
   journal= "Annual Review of Condensed Matter Physics",
   year = "2017",
   volume = "8",
   number = "Volume 8, 2017",
   pages = "337-354",
   doi = "https://doi.org/10.1146/annurev-conmatphys-031016-025458",
   url = "https://www.annualreviews.org/content/journals/10.1146/annurev-conmatphys-031016-025458",
   publisher = "Annual Reviews",
   issn = "1947-5462",
   type = "Journal Article",
  }

@article{CHENM2020,
title = {Recent Advances in Topological Quantum Materials by Angle-Resolved Photoemission Spectroscopy},
journal = {Matter},
volume = {3},
number = {4},
pages = {1114-1141},
year = {2020},
issn = {2590-2385},
doi = {https://doi.org/10.1016/j.matt.2020.07.007},
url = {https://www.sciencedirect.com/science/article/pii/S2590238520303659},
author = {Yujie Chen and Xu Gu and Yiwei Li and Xian Du and Lexian Yang and Yulin Chen}
}

@Article{LvNRP2019,
author={Lv, Baiqing
and Qian, Tian
and Ding, Hong},
title={Angle-resolved photoemission spectroscopy and its application to topological materials},
journal={Nature Reviews Physics},
year={2019},
month={Oct},
day={01},
volume={1},
number={10},
pages={609-626},
issn={2522-5820},
doi={10.1038/s42254-019-0088-5},
url={https://doi.org/10.1038/s42254-019-0088-5}
}

@Article{NeupaneNC2014,
author={Neupane, Madhab
and Xu, Su-Yang
and Sankar, Raman
and Alidoust, Nasser
and Bian, Guang
and Liu, Chang
and Belopolski, Ilya
and Chang, Tay-Rong
and Jeng, Horng-Tay
and Lin, Hsin
and Bansil, Arun
and Chou, Fangcheng
and Hasan, M. Zahid},
title={Observation of a three-dimensional topological {D}irac semimetal phase in high-mobility {Cd}$_3${As}$_2$},
journal={Nature Communications},
year={2014},
month={May},
day={07},
volume={5},
number={1},
pages={3786},
issn={2041-1723},
doi={10.1038/ncomms4786},
url={https://doi.org/10.1038/ncomms4786}
}

@Article{DengNP2016,
author={Deng, Ke
and Wan, Guoliang
and Deng, Peng
and Zhang, Kenan
and Ding, Shijie
and Wang, Eryin
and Yan, Mingzhe
and Huang, Huaqing
and Zhang, Hongyun
and Xu, Zhilin
and Denlinger, Jonathan
and Fedorov, Alexei
and Yang, Haitao
and Duan, Wenhui
and Yao, Hong
and Wu, Yang
and Fan, Shoushan
and Zhang, Haijun
and Chen, Xi
and Zhou, Shuyun},
title={Experimental observation of topological {F}ermi arcs in type-{II} {W}eyl semimetal {MoTe}$_2$},
journal={Nature Physics},
year={2016},
month={Dec},
day={01},
volume={12},
number={12},
pages={1105-1110},
issn={1745-2481},
doi={10.1038/nphys3871},
url={https://doi.org/10.1038/nphys3871}
}

@article{XuS2015,
author = {Su-Yang Xu  and Ilya Belopolski  and Nasser Alidoust  and Madhab Neupane  and Guang Bian  and Chenglong Zhang  and Raman Sankar  and Guoqing Chang  and Zhujun Yuan  and Chi-Cheng Lee  and Shin-Ming Huang  and Hao Zheng  and Jie Ma  and Daniel S. Sanchez  and BaoKai Wang  and Arun Bansil  and Fangcheng Chou  and Pavel P. Shibayev  and Hsin Lin  and Shuang Jia  and M. Zahid Hasan },
title = {Discovery of a {W}eyl fermion semimetal and topological {F}ermi arcs},
journal = {Science},
volume = {349},
number = {6248},
pages = {613-617},
year = {2015},
doi = {10.1126/science.aaa9297},
URL = {https://www.science.org/doi/abs/10.1126/science.aaa9297}}

@article{WengPRX2015,
  title = {{W}eyl Semimetal Phase in Noncentrosymmetric Transition-Metal Monophosphides},
  author = {Weng, Hongming and Fang, Chen and Fang, Zhong and Bernevig, B. Andrei and Dai, Xi},
  journal = {Phys. Rev. X},
  volume = {5},
  issue = {1},
  pages = {011029},
  numpages = {10},
  year = {2015},
  month = {Mar},
  publisher = {American Physical Society},
  doi = {10.1103/PhysRevX.5.011029},
  url = {https://link.aps.org/doi/10.1103/PhysRevX.5.011029}
}

@Article{YangP2025,
title = {Applications of electrical transport measurements in the study of topological quantum materials},
journal = {Physics},
volume = {54},
number = {5},
pages = {320-331},
year = {2025},
issn = {0379-4148},
doi = {10.7693/wl20250503},	
url = {https://wuli.iphy.ac.cn/en/article/doi/10.7693/wl20250503},
author = {YANG Yang and YANG Fan}
}

@article{HePRL2014,
  title = {Quantum Transport Evidence for the Three-Dimensional {D}irac Semimetal Phase in {Cd}$_3${As}$_2$},
  author = {He, L. P. and Hong, X. C. and Dong, J. K. and Pan, J. and Zhang, Z. and Zhang, J. and Li, S. Y.},
  journal = {Phys. Rev. Lett.},
  volume = {113},
  issue = {24},
  pages = {246402},
  numpages = {5},
  year = {2014},
  month = {Dec},
  publisher = {American Physical Society},
  doi = {10.1103/PhysRevLett.113.246402},
  url = {https://link.aps.org/doi/10.1103/PhysRevLett.113.246402}
}

@article{HuangPRX2015,
  title = {Observation of the Chiral-Anomaly-Induced Negative Magnetoresistance in {3D} {W}eyl Semimetal {TaAs}},
  author = {Huang, Xiaochun and Zhao, Lingxiao and Long, Yujia and Wang, Peipei and Chen, Dong and Yang, Zhanhai and Liang, Hui and Xue, Mianqi and Weng, Hongming and Fang, Zhong and Dai, Xi and Chen, Genfu},
  journal = {Phys. Rev. X},
  volume = {5},
  issue = {3},
  pages = {031023},
  numpages = {9},
  year = {2015},
  month = {Aug},
  publisher = {American Physical Society},
  doi = {10.1103/PhysRevX.5.031023},
  url = {https://link.aps.org/doi/10.1103/PhysRevX.5.031023}
}

@article{LiangPRX2018,
  title = {Experimental Tests of the Chiral Anomaly Magnetoresistance in the {{D}irac-{W}eyl} Semimetals {${\mathrm{Na}}_{3}\mathrm{Bi}$} and {GdPtBi}},
  author = {Liang, Sihang and Lin, Jingjing and Kushwaha, Satya and Xing, Jie and Ni, Ni and Cava, R. J. and Ong, N. P.},
  journal = {Phys. Rev. X},
  volume = {8},
  issue = {3},
  pages = {031002},
  numpages = {13},
  year = {2018},
  month = {Jul},
  publisher = {American Physical Society},
  doi = {10.1103/PhysRevX.8.031002},
  url = {https://link.aps.org/doi/10.1103/PhysRevX.8.031002}
}

@Article{ZhangN2019,
author={Zhang, Cheng
and Zhang, Yi
and Yuan, Xiang
and Lu, Shiheng
and Zhang, Jinglei
and Narayan, Awadhesh
and Liu, Yanwen
and Zhang, Huiqin
and Ni, Zhuoliang
and Liu, Ran
and Choi, Eun Sang
and Suslov, Alexey
and Sanvito, Stefano
and Pi, Li
and Lu, Hai-Zhou
and Potter, Andrew C.
and Xiu, Faxian},
title={Quantum Hall effect based on {W}eyl orbits in {Cd}$_3${As}$_2$},
journal={Nature},
year={2019},
month={Jan},
day={01},
volume={565},
number={7739},
pages={331-336},
issn={1476-4687},
doi={10.1038/s41586-018-0798-3},
url={https://doi.org/10.1038/s41586-018-0798-3}
}

@Article{UchidaNC2017,
author={Uchida, Masaki
and Nakazawa, Yusuke
and Nishihaya, Shinichi
and Akiba, Kazuto
and Kriener, Markus
and Kozuka, Yusuke
and Miyake, Atsushi
and Taguchi, Yasujiro
and Tokunaga, Masashi
and Nagaosa, Naoto
and Tokura, Yoshinori
and Kawasaki, Masashi},
title={Quantum Hall states observed in thin films of {D}irac semimetal {Cd}$_3${As}$_2$},
journal={Nature Communications},
year={2017},
month={Dec},
day={22},
volume={8},
number={1},
pages={2274},
issn={2041-1723},
doi={10.1038/s41467-017-02423-1},
url={https://doi.org/10.1038/s41467-017-02423-1}
}

@article{ShenPQE2024,
title = {Nonlinear photocurrent in quantum materials for broadband photodetection},
journal = {Progress in Quantum Electronics},
volume = {97},
pages = {100535},
year = {2024},
issn = {0079-6727},
doi = {https://doi.org/10.1016/j.pquantelec.2024.100535},
url = {https://www.sciencedirect.com/science/article/pii/S0079672724000387},
author = {Yulin Shen and Louis Primeau and Jiangxu Li and Tuan-Dung Nguyen and David Mandrus and Yuxuan Cosmi Lin and Yang Zhang}
}

@Article{ZuberFoP2021,
author={Zuber, Jack W.
and Zhang, Chao},
title={Nonlinear effects in topological materials},
journal={Frontiers of Optoelectronics},
year={2021},
month={Mar},
day={01},
volume={14},
number={1},
pages={99-109},
issn={2095-2767},
doi={10.1007/s12200-020-1088-x},
url={https://doi.org/10.1007/s12200-020-1088-x}
}

@Article{GorbarLTP2018,
author={Gorbar, E. V.
and Miransky, V. A.
and Shovkovy, I. A.
and Sukhachov, P. O.},
title={Anomalous transport properties of {D}irac and {W}eyl semimetals},
journal={Low Temperature Physics},
year={2018},
month={Jun},
day={01},
volume={44},
number={6},
pages={487-505},
issn={1063-777X},
doi={10.1063/1.5037551},
url={https://doi.org/10.1063/1.5037551}
}

@Article{WangAPS2023,
author={Wang, Huan-Wen
and Fu, Bo
and Shen, Shun-Qing},
title={Recent progress of transport theory in {D}irac quantum materials},
journal={Acta Physica Sinica},
year={2023},
volume={72},
number={17},
pages={177303-1-177303-22},
doi={10.7498/aps.72.20230672},
url={https://wulixb.iphy.ac.cn/en/article/doi/10.7498/aps.72.20230672},
url={https://doi.org/10.7498/aps.72.20230672}
}

@article{WangPRB2020,
  title = {Nonlinear current response of {W}eyl semimetals in the ultraquantum regime},
  author = {Wang, Zhigang and Fu, Zhenguo and Zhang, Ping and Zhao, Xian-Geng and Zhang, Wei},
  journal = {Phys. Rev. B},
  volume = {101},
  issue = {24},
  pages = {245313},
  numpages = {7},
  year = {2020},
  month = {Jun},
  publisher = {American Physical Society},
  doi = {10.1103/PhysRevB.101.245313},
  url = {https://link.aps.org/doi/10.1103/PhysRevB.101.245313}
}

@article{Young_PRL2012,
  title = {{D}irac Semimetal in Three Dimensions},
  author = {Young, S. M. and Zaheer, S. and Teo, J. C. Y. and Kane, C. L. and Mele, E. J. and Rappe, A. M.},
  journal = {Phys. Rev. Lett.},
  volume = {108},
  issue = {14},
  pages = {140405},
  numpages = {5},
  year = {2012},
  month = {Apr},
  publisher = {American Physical Society},
  doi = {10.1103/PhysRevLett.108.140405},
  url = {https://link.aps.org/doi/10.1103/PhysRevLett.108.140405}
}

@article{Jo_NatC2021,
  author    = {Jo, Na Hyun and Wu, Yun and Trevisan, Thaís V. and Wang, Lin-Lin and Lee, Kyungchan and Kuthanazhi, Brinda and Schrunk, Benjamin and Bud'ko, S. L. and Canfield, P. C. and Orth, P. P. and Kaminski, Adam},
  title     = {Visualizing band selective enhancement of quasiparticle lifetime in a metallic ferromagnet},
  journal   = {Nature Communications},
  year      = {2021},
  volume    = {12},
  number    = {1},
  pages     = {7169},
  month     = dec,
  doi       = {10.1038/s41467-021-27277-6},
  url       = {https://doi.org/10.1038/s41467-021-27277-6},
  issn      = {2041-1723}
}

@article{Wang_PRB2019,
  title = {Single pair of {W}eyl fermions in the half-metallic semimetal $\mathrm{EuC}{\mathrm{d}}_{2}\mathrm{A}{\mathrm{s}}_{2}$},
  author = {Wang, Lin-Lin and Jo, Na Hyun and Kuthanazhi, Brinda and Wu, Yun and McQueeney, Robert J. and Kaminski, Adam and Canfield, Paul C.},
  journal = {Phys. Rev. B},
  volume = {99},
  issue = {24},
  pages = {245147},
  numpages = {9},
  year = {2019},
  month = {Jun},
  publisher = {American Physical Society},
  doi = {10.1103/PhysRevB.99.245147},
  url = {https://link.aps.org/doi/10.1103/PhysRevB.99.245147}
}

@article{Sodemann_PRL2015,
  title = {Quantum Nonlinear Hall Effect Induced by Berry Curvature Dipole in Time-Reversal Invariant Materials},
  author = {Sodemann, Inti and Fu, Liang},
  journal = {Phys. Rev. Lett.},
  volume = {115},
  issue = {21},
  pages = {216806},
  numpages = {5},
  year = {2015},
  month = {Nov},
  publisher = {American Physical Society},
  doi = {10.1103/PhysRevLett.115.216806},
  url = {https://link.aps.org/doi/10.1103/PhysRevLett.115.216806}
}

@article{deJuan_NatC2017,
  author    = {de Juan, Fernando and Grushin, Adolfo G. and Morimoto, Takahiro and Moore, Joel E.},
  title     = {Quantized circular photogalvanic effect in {W}eyl semimetals},
  journal   = {Nature Communications},
  year      = {2017},
  volume    = {8},
  number    = {1},
  pages     = {15995},
  month     = jul,
  doi       = {10.1038/ncomms15995},
  url       = {https://doi.org/10.1038/ncomms15995},
  issn      = {2041-1723}
}

@article{Ma_NatPhys2017,
  author    = {Ma, Qiong and Xu, Su-Yang and Chan, Ching-Kit and Zhang, Cheng-Long and Chang, Guoqing and Lin, Yuxuan and Xie, Weiwei and Palacios, Tomás and Lin, Hsin and Jia, Shuang and Lee, Patrick A. and Jarillo-Herrero, Pablo and Gedik, Nuh},
  title     = {Direct optical detection of {W}eyl fermion chirality in a topological semimetal},
  journal   = {Nature Physics},
  year      = {2017},
  volume    = {13},
  number    = {9},
  pages     = {842--847},
  month     = sep,
  doi       = {10.1038/nphys4146},
  url       = {https://doi.org/10.1038/nphys4146},
  issn      = {1745-2481}
}

@article{Wu_NatPhys2017,
  author    = {Wu, Liang and Patankar, S. and Morimoto, T. and Nair, N. L. and Thewalt, E. and Little, A. and Analytis, J. G. and Moore, J. E. and Orenstein, J.},
  title     = {Giant anisotropic nonlinear optical response in transition metal monopnictide {W}eyl semimetals},
  journal   = {Nature Physics},
  year      = {2017},
  volume    = {13},
  number    = {4},
  pages     = {350--355},
  month     = apr,
  doi       = {10.1038/nphys3969},
  url       = {https://doi.org/10.1038/nphys3969},
  issn      = {1745-2481}
}

@article{Liu_PRB2025,
  title = {Theoretical investigations of high harmonic generation in the {W}eyl semimetal {WP}$_{2}$},
  author = {Liu, Xiulan and Geng, Lei and Kong, Xiao-Shuang and Zhang, Jianing and Fang, Yong-Kang and Peng, Liang-You},
  journal = {Phys. Rev. B},
  volume = {111},
  issue = {18},
  pages = {184314},
  numpages = {12},
  year = {2025},
  month = {May},
  publisher = {American Physical Society},
  doi = {10.1103/PhysRevB.111.184314},
  url = {https://link.aps.org/doi/10.1103/PhysRevB.111.184314}
}

@article{Lv_NatC2021,
  author    = {Lv, Yang-Yang and Xu, Jinlong and Han, Shuang and Zhang, Chi and Han, Yadong and Zhou, Jian and Yao, Shu-Hua and Liu, Xiao-Ping and Lu, Ming-Hui and Weng, Hongming and Xie, Zhenda and Chen, Y. B. and Hu, Jianbo and Chen, Yan-Feng and Zhu, Shining},
  title     = {High-harmonic generation in {W}eyl semimetal $\beta$-{WP}$_2$ crystals},
  journal   = {Nature Communications},
  year      = {2021},
  volume    = {12},
  number    = {1},
  pages     = {6437},
  month     = nov,
  doi       = {10.1038/s41467-021-26766-y},
  url       = {https://doi.org/10.1038/s41467-021-26766-y},
  issn      = {2041-1723}
}

@Article{Xin_AcPS2021,
title = {Circular photogalvanic effect},
journal = {Acta Physica Sinica},
volume = {70},
number = {13},
pages = {138501-1-138501-17},
year = {2021},
issn = {1000-3290},
doi = {10.7498/aps.70.20210498},	
url = {https://wulixb.iphy.ac.cn/en/article/doi/10.7498/aps.70.20210498},
author = {Su Xin and Huang Tian-Ye and Wang Jun-Zhuan and Liu Yuan and Zheng You-Liao and Shi Yi and Wang Xiao-Mu}
}

@incollection{BOYD20081,
title = {Chapter 1 - The Nonlinear Optical Susceptibility},
booktitle = {Nonlinear Optics (Third Edition)},
publisher = {Academic Press},
edition = {Third Edition},
address = {Burlington},
pages = {1-67},
year = {2008},
isbn = {978-0-12-369470-6},
doi = {https://doi.org/10.1016/B978-0-12-369470-6.00001-0},
url = {https://www.sciencedirect.com/science/article/pii/B9780123694706000010},
author = {Robert W. Boyd}
}

@article{Zhang_NatP2025,
  author    = {Zhang, Ye and Xu, David D. and Tanriover, Ibrahim and Zhou, Wenjie and Li, Yuanwei and López-Arteaga, Rafael and Aydin, Koray and Mirkin, Chad A.},
  title     = {Nonlinear optical colloidal metacrystals},
  journal   = {Nature Photonics},
  year      = {2025},
  volume    = {19},
  number    = {1},
  pages     = {20--27},
  month     = jan,
  doi       = {10.1038/s41566-024-01558-0},
  url       = {https://doi.org/10.1038/s41566-024-01558-0},
  issn      = {1749-4893}
}

@article{Salawu_NatC2025,
  author    = {Salawu, Yusuff Adeyemi and Choun, Yoonseok and Yoo, Seung-Jong and Sasaki, Minoru and Kim, Ki-Seok and Kim, Heon-Jung},
  title     = {Observation of symmetry-forbidden rectification in an inversion-symmetric {W}eyl metal},
  journal   = {Nature Communications},
  year      = {2025},
  volume    = {16},
  number    = {1},
  pages     = {4153},
  month     = may,
  doi       = {10.1038/s41467-025-59394-x},
  url       = {https://doi.org/10.1038/s41467-025-59394-x},
  issn      = {2041-1723}
}

@article{Du_NatC2019,
  author    = {Du, Z. Z. and Wang, C. M. and Li, Shuai and Lu, Hai-Zhou and Xie, X. C.},
  title     = {Disorder-induced nonlinear Hall effect with time-reversal symmetry},
  journal   = {Nature Communications},
  year      = {2019},
  volume    = {10},
  number    = {1},
  pages     = {3047},
  month     = jul,
  doi       = {10.1038/s41467-019-10941-3},
  url       = {https://doi.org/10.1038/s41467-019-10941-3},
  issn      = {2041-1723}
}

@article{Suo_ACSP2026,
    author = {Suo, Peng and Geng, Long and Yang, Yunkun and Wang, Chen and Zhang, Yidan and Lin, Xian and Zhang, Chao and Xiu, Faxian and Ma, Guohong},
    title = {Unlocking and
Controlling Efficient Second-Order Nonlinear
Terahertz Photocurrents in Centrosymmetric {D}irac Semimetals},
    journal = {ACS Photonics},
    volume = {13},
    number = {2},
    pages = {582-591},
    year = {2026},
    month = {01},
    issn = {2330-4022},
    doi = {10.1021/acsphotonics.5c02677},
    url = {https://doi.org/10.1021/acsphotonics.5c02677},
    eprint = {https://pubs.acs.org/apchd5/article-pdf/13/2/582/63485948/ph5c02677.pdf},
}

@article{Du_NatRP2021,
  author    = {Du, Z. Z. and Lu, Hai-Zhou and Xie, X. C.},
  title     = {Nonlinear Hall effects},
  journal   = {Nature Reviews Physics},
  year      = {2021},
  volume    = {3},
  number    = {11},
  pages     = {744--752},
  month     = nov,
  doi       = {10.1038/s42254-021-00359-6},
  url       = {https://doi.org/10.1038/s42254-021-00359-6},
  issn      = {2522-5820}
}

@article{Xiao_PRB2019,
  title = {Theory of nonlinear Hall effects: Modified semiclassics from quantum kinetics},
  author = {Xiao, Cong and Du, Z. Z. and Niu, Qian},
  journal = {Phys. Rev. B},
  volume = {100},
  issue = {16},
  pages = {165422},
  numpages = {7},
  year = {2019},
  month = {Oct},
  publisher = {American Physical Society},
  doi = {10.1103/PhysRevB.100.165422},
  url = {https://link.aps.org/doi/10.1103/PhysRevB.100.165422}
}

@article{Ma_PRB2025,
  title = {Quantum kinetic theory of the semiclassical side jump, skew scattering, and longitudinal velocity},
  author = {Ma, Da and Zhang, Zhi-Fan and Jiang, Hua and Xie, X. C.},
  journal = {Phys. Rev. B},
  volume = {112},
  issue = {4},
  pages = {045136},
  numpages = {13},
  year = {2025},
  month = {Jul},
  publisher = {American Physical Society},
  doi = {10.1103/rgr5-yzy2},
  url = {https://link.aps.org/doi/10.1103/rgr5-yzy2}
}

@article{3r17-kfy7,
  title = {Thermoelectric transport in graphene under strain fields modeled by Dirac oscillators},
  author = {Ca\~nas, Juan A. and Bonilla, Daniel A. and Mart\'{\i}n-Ruiz, A.},
  journal = {Phys. Rev. B},
  volume = {112},
  issue = {10},
  pages = {104206},
  numpages = {14},
  year = {2025},
  month = {Sep},
  publisher = {American Physical Society},
  doi = {10.1103/3r17-kfy7},
}

@article{canas2026charge,
  title={Charge and energy transport in graphene with smooth finite-range disorder},
  author={Ca{\~n}as, Juan A and Bonilla, Daniel A and P{\'e}rez-Pedraza, JC and Mart{\'\i}n-Ruiz, A},
  journal={Physica B: Condensed Matter},
  pages={418431},
  year={2026},
  volume = {729},
  issn = {0921-4526},
  publisher={Elsevier},
  doi = {https://doi.org/10.1016/j.physb.2026.418431},
}

@article{Sinitsyn_2008,
doi = {10.1088/0953-8984/20/02/023201},
url = {https://doi.org/10.1088/0953-8984/20/02/023201},
year = {2007},
month = {dec},
publisher = {},
volume = {20},
number = {2},
pages = {023201},
author = {Sinitsyn, N A},
title = {Semiclassical theories of the anomalous Hall effect},
journal = {Journal of Physics: Condensed Matter}
}

@article{Zyuzin_PRB2012,
  title = {Weyl semimetal with broken time reversal and inversion symmetries},
  author = {Zyuzin, A. A. and Wu, Si and Burkov, A. A.},
  journal = {Phys. Rev. B},
  volume = {85},
  issue = {16},
  pages = {165110},
  numpages = {9},
  year = {2012},
  month = {Apr},
  publisher = {American Physical Society},
  doi = {10.1103/PhysRevB.85.165110},
  url = {https://link.aps.org/doi/10.1103/PhysRevB.85.165110}
}

@article{Chang_PRB2015,
  title = {{RKKY interaction of magnetic impurities in Dirac and Weyl semimetals}},
  author = {Chang, Hao-Ran and Zhou, Jianhui and Wang, Shi-Xiong and Shan, Wen-Yu and Xiao, Di},
  journal = {Phys. Rev. B},
  volume = {92},
  issue = {24},
  pages = {241103(R)},
  numpages = {5},
  year = {2015},
  month = {Dec},
  publisher = {American Physical Society},
  doi = {10.1103/PhysRevB.92.241103},
  url = {https://link.aps.org/doi/10.1103/PhysRevB.92.241103}
}

@article{DasSarma_PRB2015,
  title = {Carrier screening, transport, and relaxation in three-dimensional Dirac semimetals},
  author = {Das Sarma, S. and Hwang, E. H. and Min, Hongki},
  journal = {Phys. Rev. B},
  volume = {91},
  issue = {3},
  pages = {035201},
  numpages = {11},
  year = {2015},
  month = {Jan},
  publisher = {American Physical Society},
  doi = {10.1103/PhysRevB.91.035201},
  url = {https://link.aps.org/doi/10.1103/PhysRevB.91.035201}
}

@article{Ominato_PRB2014,
  title = {Quantum transport in a three-dimensional Weyl electron system},
  author = {Ominato, Yuya and Koshino, Mikito},
  journal = {Phys. Rev. B},
  volume = {89},
  issue = {5},
  pages = {054202},
  numpages = {8},
  year = {2014},
  month = {Feb},
  publisher = {American Physical Society},
  doi = {10.1103/PhysRevB.89.054202},
  url = {https://link.aps.org/doi/10.1103/PhysRevB.89.054202}
}

@article{Ominato_PRB2015,
  title = {Quantum transport in three-dimensional Weyl electron system in the presence of charged impurity scattering},
  author = {Ominato, Yuya and Koshino, Mikito},
  journal = {Phys. Rev. B},
  volume = {91},
  issue = {3},
  pages = {035202},
  numpages = {9},
  year = {2015},
  month = {Jan},
  publisher = {American Physical Society},
  doi = {10.1103/PhysRevB.91.035202},
  url = {https://link.aps.org/doi/10.1103/PhysRevB.91.035202}
}

@article{Hosseini_PRB2015,
  title = {{Ruderman-Kittel-Kasuya-Yosida interaction in Weyl semimetals}},
  author = {Hosseini, Mir Vahid and Askari, Mehdi},
  journal = {Phys. Rev. B},
  volume = {92},
  issue = {22},
  pages = {224435},
  numpages = {6},
  year = {2015},
  month = {Dec},
  publisher = {American Physical Society},
  doi = {10.1103/PhysRevB.92.224435},
  url = {https://link.aps.org/doi/10.1103/PhysRevB.92.224435}
}

@article{Principi_PRB2015,
  title = {{Kondo effect and non-Fermi-liquid behavior in Dirac and Weyl semimetals}},
  author = {Principi, Alessandro and Vignale, Giovanni and Rossi, E.},
  journal = {Phys. Rev. B},
  volume = {92},
  issue = {4},
  pages = {041107(R)},
  numpages = {5},
  year = {2015},
  month = {Jul},
  publisher = {American Physical Society},
  doi = {10.1103/PhysRevB.92.041107},
  url = {https://link.aps.org/doi/10.1103/PhysRevB.92.041107}
}

@article{Morimoto_PRB2016,
  title = {Semiclassical theory of nonlinear magneto-optical responses with applications to topological Dirac/Weyl semimetals},
  author = {Morimoto, Takahiro and Zhong, Shudan and Orenstein, Joseph and Moore, Joel E.},
  journal = {Phys. Rev. B},
  volume = {94},
  issue = {24},
  pages = {245121},
  numpages = {15},
  year = {2016},
  month = {Dec},
  publisher = {American Physical Society},
  doi = {10.1103/PhysRevB.94.245121},
  url = {https://link.aps.org/doi/10.1103/PhysRevB.94.245121}
}

@article{Nagaosa_RMP2010,
  title = {{Anomalous Hall effect}},
  author = {Nagaosa, Naoto and Sinova, Jairo and Onoda, Shigeki and MacDonald, A. H. and Ong, N. P.},
  journal = {Rev. Mod. Phys.},
  volume = {82},
  issue = {2},
  pages = {1539--1592},
  numpages = {0},
  year = {2010},
  month = {May},
  publisher = {American Physical Society},
  doi = {10.1103/RevModPhys.82.1539},
  url = {https://link.aps.org/doi/10.1103/RevModPhys.82.1539}
}

@article{Golub_PRB2018,
  title = {{Circular and magnetoinduced photocurrents in Weyl semimetals}},
  author = {Golub, L. E. and Ivchenko, E. L.},
  journal = {Phys. Rev. B},
  volume = {98},
  issue = {7},
  pages = {075305},
  numpages = {13},
  year = {2018},
  month = {Aug},
  publisher = {American Physical Society},
  doi = {10.1103/PhysRevB.98.075305},
  url = {https://link.aps.org/doi/10.1103/PhysRevB.98.075305}
}
\end{document}